\documentclass[aps,showpacs,superscriptaddress,nofootinbib]{revtex4}
\usepackage[dvips]{graphicx}
\usepackage{epsfig,rotating}
\usepackage{amsmath,amssymb}
\numberwithin{equation}{section}

\newcommand{\be}{\begin{equation}}
\newcommand{\ee}{\end{equation}}
\newcommand{\bea}{\begin{eqnarray}}
\newcommand{\eea}{\end{eqnarray}}
\newcommand{\vk}{\vec k}
\newcommand{\vx}{\vec x}

\begin{document}
\title{Oscillations and evolution of a hot and dense gas of flavor neutrinos:\\a quantum field theory study.}
\author{D. Boyanovsky}
\email{boyan@pitt.edu} \affiliation{Department of Physics and
Astronomy, University of Pittsburgh, Pittsburgh, Pennsylvania
15260, USA}
\author{C.M. Ho}
\email{cmho@phyast.pitt.edu} \affiliation{Department of Physics
and Astronomy, University of Pittsburgh, Pittsburgh, Pennsylvania
15260, USA}

\date{\today}

\begin{abstract}
We study the time evolution of the distribution functions for hot
and or degenerate gases of two flavors of Dirac neutrinos as a
result of flavor mixing and dephasing. This is achieved by
obtaining the time evolution of the flavor density matrix directly
from quantum field theory at finite temperature and density. The
time evolution features a rich hierarchy of  scales which are
widely separated in the nearly degenerate or relativistic cases
and originate in interference phenomena between particle and
antiparticle states. In the degenerate case the flavor asymmetry
$\Delta N(t)$ relaxes to the asymptotic limit $\Delta
N(\infty)=\Delta N(0)\cos^2(2\theta)$ via dephasing resulting from
the oscillations between flavor modes that are not Pauli blocked,
with a power law $1/t$ for $t>t_s \approx 2 k_F/\Delta M^2$. $k_F$
is the largest of the Fermi momenta. The distribution function for
flavor neutrinos and antineutrinos as well as off-diagonal
densities are obtained. Flavor particle-antiparticle pairs are
produced by mixing and oscillations with typical momentum $k\sim
\bar{M}$ the average mass of the neutrinos. An effective field
theory description emerges on long time scales in which the
Heisenberg operators obey a Bloch-type equation of motion valid in
the relativistic and nearly degenerate cases. We find the
non-equilibrium propagators and correlation functions in this
effective theory and discuss its regime of validity  as well as
the potential corrections.
\end{abstract}

 \maketitle

\section{Introduction}\label{sec:intro}
Neutrinos are the bridge between particle physics, astrophysics,
cosmology and nuclear
physics\cite{book1,book2,book3,raffelt,pantaleone,dolgov}, and
after almost four decades of the prescient suggestion that
neutrinos may oscillate\cite{pontecorvo,bilponte}, a wealth of
experimental data confirms that neutrinos are massive and that
different flavors mix and
oscillate\cite{giunti,smirnov,bilenky,haxton,grimus}. Neutrino
masses and mixing decidedly points to  new physics beyond the
standard model and profoundly impacts on the physics, astrophysics
and cosmology of neutrinos. Neutrino oscillations in matter may
provide an explanation of the solar neutrino problem by the
resonant conversion of flavor neutrinos in the medium, namely the
MSW effect\cite{MSWI,MSWII} (for recent reviews
see\cite{book1}-\cite{Beuthe}). The dynamical aspects of neutrino
oscillations in extreme conditions of temperature and density play
an important role in Big Bang Nucleosynthesis (BBN) and in the
lepton asymmetry in the early Universe\cite{kirilova} (for a
recent review see\cite{dolgov}) as well as in the physics of core
collapse supernovae and the formation, evolution and cooling of
neutron stars\cite{prakash,reddy,yakovlev}. The study of the
dynamical evolution of a hot and/or dense gas of neutrinos that
include mixing as well as collisions has been and continues to be
the subject of much attention in the literature. Neutrino mixing
and oscillations introduce a novel aspect in the description of
flavor equilibration, since the weak interactions involve flavor
(weak) eigenstates while time evolution is described in terms of
mass eigenstates. Therefore in a dense and/or hot medium where
neutrino interactions cannot be neglected collisional processes
must be studied on the same footing as the dynamics of
oscillations. Furthermore in a dense background of neutrinos such
as is the case in the early Universe during the time relevant for
BBN or during the time scale of neutrino trapping in a
protoneutron star, the neutral current interaction leads to a
contribution to the neutrino self-energy from forward scattering
off the neutrino background akin to the contribution from the
electron plasma that leads to MSW resonance
enhancement\cite{MSWI}. In dense neutrino gases, this self-energy
contribution leads to a non-linear problem for the evolution of a
given neutrino interacting with the neutrino background.

The dynamics of neutrino oscillations was originally studied in
terms of Bloch-type equations  akin to the equation of motion for
a spin in a magnetic field\cite{stodolsky,MSWI,book1,book2,book3}
which are generally valid for single particle descriptions in the
relativistic limit. For the case of \textit{single particle
states} this equation of motion for neutrino oscillations was
derived from the underlying field theory in the relativistic
limit\cite{mannheim,pantaleone}. This formulation of the dynamics
of oscillations of single particle states was extended to a
kinetic description of oscillations and mixing in a
medium\cite{sigl2,enqvist,mckellar}.  The resulting equations in
principle include the effects of collisions as well as the
non-linearities arising from neutrino forward scattering off a
neutrino background. They have been implemented to study the
evolution of the neutrino distribution functions in the early
Universe\cite{dolgov,barbieri,samuel,lunar1,wong,beacom}, in
supernovae\cite{pantaleone2,fuller,sigl3} as well as to study the
relic neutrino asymmetry\cite{bell}. Novel fascinating
self-synchronization phenomena emerges as a consequence of the
non-linearities in a neutrino background with potential
implications on CP (and baryon) asymmetry in the early
Universe\cite{samuel,dolgov}.

An alternative quantum field theory treatment of neutrinos in the
medium used the ingredients of thermal field
theory\cite{notzold,nieves} combined with a self-consistent
treatment in the case of a neutrino background\cite{dolivo}. Since
the main method in this approach relies on the equilibrium
description of thermal field theory, there is an underlying
assumption that the neutrino background is nearly in equilibrium.

More recently the validity of the single particle picture that
underlies the kinetic equations for neutrinos in a medium has been
critically re-examined\cite{lunardini}.

In our view, the study of neutrino oscillations and mixing in the
case of a dense and or hot neutrino background either via the set
of kinetic equations\cite{sigl2,enqvist,mckellar} or the thermal
field theory approach invoke a variety of approximations some of
which are not very clear. In the kinetic description several
approximations are involved, from neglecting interference terms
between particles and antiparticles by restricting the
Hamiltonian\cite{sigl2} to some time averaging and restriction to
single particle evolution\cite{mckellar}. Some of these
approximations motivated the study of ref.\cite{lunardini}.

A full quantum field theory treatment of  neutrino mixing reveals
a more complex picture of oscillations beyond that of the single
particle description\cite{blasone,fujii,ji}. The authors of these
references pointed out that a careful treatment of the Fock
representation of flavor states leads to novel contributions to
the oscillation formula even for single particle states. While it
has been argued recently that  Fock states of flavor neutrinos may
not be relevant for S-matrix processes\cite{giunti2} a quantum
statistical mechanics of dense and or hot flavor neutrino gases
must necessarily rely on the Fock representation (occupation
number) for flavor neutrino states.

A quantum statistical description of a dense and or hot gas of
\emph{flavor neutrinos} requires the notion of an occupation
number which inevitably implies a description in terms of Fock
flavor states. Furthermore a chemical potential associated with a
flavor neutrino is a variable conjugate to the number of these
flavor  neutrinos.

Regardless of whether the variety of approximations usually
invoked are justified for practical purposes, the study of the
dynamics of neutrino mixing and oscillations from the point of
view of quantum field theory is clearly of fundamental importance
as a prelude towards physics beyond the standard model. While
there have been studies of the quantum field theory aspects in
\emph{vacuum} we are not aware of any previous study of the
quantum field theory of mixing in a dense and or hot medium with
neutrinos.

\vspace{2mm}

\textbf{The goal of this study:}

\vspace{2mm}

In the presence of flavor mixing, individual flavor number is not
conserved and a density matrix that is diagonal in the flavor Fock
basis will evolve in time and develop off diagonal elements.

Hence time evolution of a dense or hot neutrino gas has to be
studied as a quantum mechanical initial value problem: an initial
density matrix which is diagonal in the flavor basis is evolved in
time with the full Hamiltonian with flavor mixing. In this article
we focus on studying precisely the time evolution of a dense or
hot flavor neutrino gas in the simplest case of \textbf{free field
theory}. Our goal is to study the evolution of an initially
prepared density matrix which is diagonal in the flavor basis and
describes a quantum  gas of flavor neutrinos at finite density or
finite temperature. We undertake the study of the dynamics in
\textbf{free field theory} as a prelude towards a complete
understanding of oscillation phenomena in weak interactions. The
first step of any systematic program must be the understanding at
the simplest level. As will be detailed below, studying the
dynamics of oscillations and mixing in a dense and/or hot medium
even at the level of free field theory reveals a wealth of subtle
and important phenomena which leads to a firmer understanding of
the validity of the various approximations as well as highlighting
the potential corrections.

The problem that we study can be stated succinctly as
follows\footnote{D.B. thanks S. Reddy for stating the question
over dinner.}: Consider that at a given initial time we have a
``box'' that contains a hot or dense gas of flavor neutrinos with
a given single particle distribution consistent with Fermi-Dirac
statistics , how does this \emph{ensemble} evolve in time?, how do
the populations of flavor neutrinos evolve in time?, how do flavor
neutrinos \emph{propagate} in the medium??.

While our ultimate goal is to study the evolution in the presence
of the weak interactions, we begin our study in this simplest free
field theory case and the case of two flavors with the following
goals in mind

\begin{itemize}
\item{To study the evolution directly from the underlying quantum
field theory \emph{without} making any approximations. This study
will clarify the nature of the various approximations invoked in
the literature and exhibit the potential corrections.  }

\item{By keeping the full evolution,  the different time scales
will emerge thus paving the way to providing a firmer
understanding of coherence effects as well as the time averaging
implied by several approximations.}

\item{A first principle derivation of kinetic equations and or
Boltzmann equations require the propagators for the
fields\cite{boyboltz} in the medium. Thus the study of the
evolution in free field theory is the starting point for a
systematic treatment of oscillations and collisions in a medium
with a neutrino background. }

\item{ As it will become clear below, the study of even the simple
free field theory case reveals a wealth of phenomena as a
consequence of flavor mixing, which to the best of our knowledge
has not been recognized and explored fully before in the case of
finite temperature and density. The full quantum field theory
treatment unambiguously reveals all the complexities associated
with flavor mixing and allows  a systematic implementation of
several approximations which clarify the regime of validity of the
single particle description and provide an understanding of the
corrections. }
\end{itemize}

\textbf{Brief summary of the results:} Our main results are
briefly summarized as follows,

\begin{itemize}
\item{The dynamics of neutrino oscillations of a dense and or hot
gas of flavor neutrinos features a hierarchy of time scales. The
fast time scales are associated with  interference effects between
particle and \emph{antiparticle} states while the slow scales
emerge from interference between particle states (or antiparticle
states) of different masses. In the nearly degenerate or
relativistic case the scales are widely separated and processes
which involve interference between particle and antiparticle
states become subdominant on the slow dynamics.  }

\item{ An initial flavor asymmetry relaxes towards an asymptotic
value $\Delta N(\infty)=\Delta N(0)\cos^2(2\theta)$ (with $\theta$
being the mixing angle) with a power law $\propto 1/t$ as a
consequence of \emph{dephasing}. Pauli blocking manifests in that
neutrinos of one flavor can only oscillate into un-occupied states
of neutrinos of different flavor and dephasing is a consequence of
oscillations between Pauli unblocked flavor states. We obtain the
explicit time evolution of the distribution functions as well as
off-diagonal correlation functions. We discuss the phenomenon of
\emph{flavor pair production} by mixing and oscillations. This is
a consequence of the overlap between particle and antiparticle
states and results in the production of pairs of flavored
neutrinos with typical momenta $k\sim \bar{M}$, the average mass
of the neutrinos. }

\item{ In the nearly degenerate case (as suggested by the recent
combined observations) or in the relativistic case as is likely to
prevail in the early universe as well as in core collapse
supernovae, the different time scales are widely separated. This
allows to establish an ``effective'' (free) field theory
description valid on the slow time scales. The equations of motion
for Heisenberg operators in this effective description are the oft
quoted Bloch-type equations, but the effective field theory also
describes the quantum fields. This effective theory allows to
construct the Feynman propagators which feature distinct
\emph{non-equilibrium} aspects and to  clearly identify the
potential corrections and is valid both in the relativistic as
well as in the nearly degenerate case. }

\end{itemize}

Our study is organized as follows: in section II the theory
corresponding to two flavors of neutrinos as well as the density
matrix that describes an initial state of flavor neutrinos is
presented. In this section we address the quantization aspects and
point out the source of subtle mixing phenomena between particles
and \emph{antiparticles}, confirming previous results in the
literature\cite{blasone}. In sections III and IV we study the
evolution of the flavor asymmetry as well as that of the
individual distribution functions focusing on the emergence of a
hierarchy of scales and extracting the asymptotic long time
dynamics as well as the phenomenon of flavor pair production via
oscillations. In section V we present the ``effective'' field
theory that describes the long-time dynamics and discuss its
regime of validity. In this section we obtain the Feynman
propagators and discuss their non-equilibrium aspects. In section
VI we discuss the regime of validity of the several approximations
as well as caveats in the formulation and present our conclusions.

\section{Neutrino mixing and flavor density matrix}

We focus our attention on the evolution of \emph{Dirac} neutrinos
postponing the case of Majorana neutrinos for further discussion
elsewhere. Furthermore, we restrict the discussion to the case of
two flavors which provides the simplest scenario. Most of the
results can be extrapolated to the case of three active flavors
including the case of sterile neutrinos, but for the subtleties
associated with CP violating phases which of course are of great
interest but will not be addressed here. We will call the flavors
the electron and muon neutrino, but the results apply more broadly
to  active-sterile oscillations.

Consider the Dirac neutrino fields with the Lagrangian density
given by

\begin{equation}
\mathcal{L} =\bar{\nu}_{e}(x) (i\!\!\not\!\partial) \nu_{e}(x) +
             \bar{\nu}_{\mu}(x) (i\!\!\not\!\partial) \nu_{\mu}(x) +
\left(%
\begin{array}{cc}
 \bar{\nu}_{e}(x) & \bar{\nu}_{\mu}(x) \\
\end{array}%
\right)
\left(%
\begin{array}{cc}
  m_{e} & m_{e \mu} \\
  m_{e \mu } & m_{\mu } \\
\end{array}%
\right)
\left(%
\begin{array}{c}
  \nu_{e}(x) \\
  \nu_{\mu}(x) \\
\end{array}%
\right), \label{L}
\end{equation}

\noindent where $m_{e\mu}$ is the mixing  and we have absorbed a
potential phase into a field redefinition. The mass matrix can be
diagonalized by introducing a rotation matrix such that

\begin{equation}
\left(%
\begin{array}{c}
  \nu_{e}(x) \\
  \nu_{\mu}(x) \\
\end{array}%
\right) =
\left(%
\begin{array}{cc}
  C & S \\
  -S & C \\
\end{array}%
\right)
\left(%
\begin{array}{c}
  \psi_{1}(x) \\
  \psi_{2}(x) \\
\end{array}%
\right), \label{RotationMatrix}
\end{equation}

where for simplicity of notation we  defined

\begin{equation}\label{angles}
C\equiv \cos\theta \; ; \;  S\equiv \sin\theta \end{equation}

\noindent where $\theta$ is the mixing angle. The diagonalized
mass matrix then reads

\begin{equation}
\left(%
\begin{array}{cc}
  M_{1} & 0 \\
  0 & M_{2} \\
\end{array}%
\right) =
\left(%
\begin{array}{cc}
  C & -S \\
  S & C \\
\end{array}%
\right)
\left(%
\begin{array}{cc}
  m_{e} & m_{e \mu} \\
  m_{e \mu } & m_{\mu } \\
\end{array}%
\right)
\left(%
\begin{array}{cc}
  C & S \\
  -S & C \\
\end{array}%
\right).
\end{equation}

In the mass eigenstate basis, the Lagrangian density becomes

\begin{equation}
\mathcal{L}=\bar{\psi}_{1}(x)(i\!\!\not\!\partial-M_{1})\psi_{1}(x)+
            \bar{\psi}_{2}(x)(i\!\!\not\!\partial-M_{2})\psi_{2}(x).
\end{equation}

In what follows, we reserve the latin label $i=1,2$ for the fields
associated with the mass eigenstates $\psi$ and the greek label
$\alpha=e,\,\mu$ for the fields associated with the flavor
eigenstates $\nu$.

Upon quantization in a volume $V$, the flavor field operators
$\nu_{\alpha}(x)$ at time $t=0$ are written as
\begin{eqnarray}\label{flavor}
&& \nu_{\alpha}(\vx)
 =\frac{1}{\sqrt{V}}\sum_{\vk}
 \nu_{\alpha}(\vk)\, e^{i\vec{k}\cdot\vec{x}} ,\nonumber\\
&& \nu_{\alpha}(\vk)= \sum_{\lambda}
 \left(
 \alpha_{\vk, \lambda}^{(\alpha)}U^{(\alpha)}_{\vk,\lambda}+
 \beta_{-\vk , \lambda}^{(\alpha) \dagger}V_{-\vk , \lambda}^{(\alpha)}
 \right)
\end{eqnarray}

\noindent where the index $\lambda$ refers to the Dirac spin index
and we have kept the same notation for the field and its spatial
Fourier transform to avoid cluttering of notation. A flavor Fock
representation is defined by choosing the spinors $U$ and $V$
respectively. In principle these spinors can be chosen to be the
positive and negative energy solutions of a Dirac equation with an
arbitrary mass, in what follows we will choose these to be
$m_{e};m_{ \mu}$, namely the masses of the flavor eigenstates in
the \emph{absence of mixing}. While we consider this to be a
physically motivated choice, it is by no means unique and
different alternatives have been discussed in the
literature\cite{blasone,fujii,ji}.

Thus the spinors $U$ and $V$ are chosen to be solutions of the
following Dirac equations

\begin{eqnarray}\label{diracflavor}
&&\gamma^0(\vec{\gamma}\cdot \vk + m_{\alpha})\,
U^{(\alpha)}_{\vk,\lambda}  =
\omega_{\alpha}(k)\, U^{(\alpha)}_{\vk,\lambda}\nonumber \\
&&\gamma^0(\vec{\gamma}\cdot \vk + m_{\alpha})\,
V^{(\alpha)}_{-\vk,\lambda}  =
 -\omega_{\alpha}(k) \, V^{(\alpha)}_{-\vk,\lambda}\\
&& \omega_{\alpha}(k) =  \sqrt{k^2+m^2_{\alpha}}
\end{eqnarray}

The Dirac spinors $U$ and $V$, are normalized as follows (no sum
over the index $\alpha$)

\begin{equation}\label{normaflavor}
U_{\vk ,\lambda}^{(\alpha)\dagger}U_{\vk ,
\lambda^{\prime}}^{(\alpha)} = V_{\vk ,
\lambda}^{(\alpha)\dagger}V_{\vk , \lambda^{\prime}}^{(\alpha)}
=\delta_{\lambda , \lambda^{\prime}}\;;\;
U_{\vk,\lambda}^{(\alpha)\dagger}V_{-\vk,\lambda^{\prime}}^{(\alpha)}
=0.
\end{equation}
 \noindent and the creation and annihilation operators $\alpha_{\vk,
 \lambda};\beta_{\vk, \lambda}$ obey the usual canonical
 anticommutation relations.

On the other hand, upon quantization the field operators
$\psi_{i}(x)$ associated with mass eigenstates at time $t=0$ are
given by

\begin{eqnarray}\label{psis}
&& \psi_{i}(\vx)
 =\frac{1}{\sqrt{V}}\sum_{k} \psi_{i}(\vk)\, e^{i\vec{k}\cdot\vec{x}} \nonumber \\
&&  \psi_{i}(\vk) = \sum_{\lambda} \left(
 a_{\vk, \lambda}^{(i)}F_{\vk, \lambda}^{(i)}+
 b_{-\vk, \lambda}^{(i) \dagger}G_{-\vk, \lambda}^{(i)}
 \right).
\end{eqnarray}

\noindent where the spinors $F,G$ are now solutions of the
following Dirac equations

\begin{eqnarray}\label{diracmass}
&&\gamma^0(\vec{\gamma}\cdot \vk + M_{i})\, F^{(i)}_{\vk,\lambda}
=
E_{i}(k)\, F^{(i)}_{\vk,\lambda}\nonumber \\
&&\gamma^0(\vec{\gamma}\cdot \vk + M_{i})\, G^{(i)}_{-\vk,\lambda}
=
 -E_{i}(k) \, G^{(i)}_{-\vk,\lambda}\\
&& E_{i}(k) =  \sqrt{k^2+M^2_{i}}
\end{eqnarray}

\noindent with the normalization conditions (no sum over the label
$i$)

\begin{equation}\label{normamass}
F_{\vk ,\lambda}^{(i)\dagger}F_{\vk , \lambda^{\prime}}^{(i)} =
G_{\vk , \lambda}^{(i)\dagger}G_{\vk , \lambda^{\prime}}^{(i)}
=\delta_{\lambda , \lambda^{\prime}}\; ; \;
F_{\vk,\lambda}^{(i)\dagger}G_{-\vk,\lambda^{\prime}}^{(i)} =0.
\end{equation}

Similarly, the operators $a$ and $b$ satisfy usual canonical
anticommutation relations.

\subsection{Hamiltonian and Charges}\label{hamchar}

The total free field Hamiltonian for mixed neutrinos in the
diagonal (mass) basis is given by

\begin{equation}\label{totalham}
H=\sum_{\vk,i} \left[\bar{\psi}_i(\vk)(\vec{\gamma}\cdot \vk +
M_{i})\psi_i(\vk)  \right]= \sum_{\vk,\lambda,i}\left(
a_{\vk,\lambda}^{(i)\dagger}a_{\vk,\lambda}^{(i)}+ b_{\vk,
\lambda}^{(i)\dagger}b_{\vk,\lambda}^{(i)} -1\right)E_{i}(k),
\end{equation}

Therefore the time evolution of the operators $a,b$  is given by

\begin{eqnarray}
a_{\vk,\lambda}^{(i)}(t)&=&a_{\vk,\lambda}^{(i)}e^{-iE_i(k)t}\nonumber \\
b_{\vk,\lambda}^{(i)}(t)&=& b_{\vk,\lambda}^{(i)}e^{-iE_i(k)t}.
\end{eqnarray}

The free field Lagrangian density (\ref{L}) is invariant under
independent phase transformations of the fields $\psi_{1,2}$,
hence the individual $U(1)$ charges

\begin{equation}\label{chargesmass}
Q_i = \int d^3 x\;
\psi^{\dagger}_i(\vx,t)\psi_i(\vx,t)=\sum_{\vk,\lambda}
\left[a_{\vk,\lambda}^{(i)\dagger}a_{\vk,\lambda}^{(i)}- b_{\vk,
\lambda}^{(i)\dagger}b_{\vk,\lambda}^{(i)} +1\right]
\end{equation}

\noindent are time independent.

The discussion that follows will focus on describing a statistical
density matrix which is \emph{diagonal} in the flavor basis and
describes a hot and or dense ensemble of flavor neutrinos. This
discussion requires the \emph{flavor} Hamiltonian which is
obtained from the Lagrangian density (\ref{L}) for vanishing
mixing $m_{e\mu}=0$, namely

\begin{equation}\label{hamflavor}
H_f=H_e+H_{\mu}=\sum_{\vk,\alpha}
\left[\bar{\nu}_{\alpha}(\vk)(\vec{\gamma}\cdot \vk +
m_{\alpha})\nu_{\alpha}(\vk) \right]=
\sum_{\vk,\lambda,\alpha}\left(
\alpha_{\vk,\lambda}^{(\alpha)\dagger}\alpha_{\vk,\lambda}^{(\alpha)}+
\beta_{\vk,
\lambda}^{(\alpha)\dagger}\beta_{\vk,\lambda}^{(\alpha)}-1
\right)\omega_{\alpha}(k),
\end{equation}

The flavor Hamiltonian above is invariant under independent phase
transformations of the flavor fields $\nu_{\alpha}$, thus the
individual flavor charges commute with $H_f$

\begin{equation}\label{chargeflavor}
q_\alpha = \int d^3 x\;
\nu^{\dagger}_{\alpha}(\vx)\nu_{\alpha}(\vx)=\sum_{\vk}\nu^{\dagger}_{\alpha}(\vk)\nu_{\alpha}(\vk)=
\sum_{\vk,\lambda}
\left[\alpha_{\vk,\lambda}^{(\alpha)\dagger}\alpha_{\vk,\lambda}^{(\alpha)}-
\beta_{\vk,
\lambda}^{(\alpha)\dagger}\beta_{\vk,\lambda}^{(\alpha)} +1\right]
\end{equation}

Using the transformation law (\ref{RotationMatrix}) between flavor
and mass eigenstates it is straightforward to find that the total
charges are the same, namely

\begin{equation}\label{totalcharge}
\sum_{i,\vk}\psi_i^{\dagger}(\vk,t)\psi_i(\vk,t)=\sum_{\alpha,\vk}\nu_{\alpha}^\dagger(\vk,t)\nu_{\alpha}(\vk,t)
\Rightarrow Q_1+Q_2 = q_e + q_{\mu}
\end{equation}

\subsection{Density matrix and time evolution}

As stated in the introduction, our focus and goal is to study the
time evolution of the distribution function of flavor neutrinos,
 at the level of free field theory at this stage. The question that
we posed in the introduction and address here is the following:
consider that at some given time the gas of flavor neutrinos and
antineutrinos are described by a quantum statistical ensemble with
a Fermi-Dirac distribution function with a fixed chemical
potential for each flavor, namely

\begin{equation}\label{flavnumb}
n^{(\alpha)}(k)= \frac{1}{e^{\beta(
\omega_{\alpha}(k)-\mu_\alpha)}+1} \; ; \; \bar{n}^{(\alpha)}(k)=
\frac{1}{e^{\beta (\omega_{\alpha}(k)+\mu_\alpha)}+1}
\end{equation}

\noindent with $\beta= 1/T$ and $\mu_{\alpha}$  the chemical
potential for each flavor.

Such an ensemble is described by a quantum statistical density
matrix which is \emph{diagonal} in the Fock space of flavor
eigenstates and is given by

\begin{equation}\label{rhotot}
\hat{\rho} = \hat{\rho}^{(e)}\otimes \hat{\rho}^{(\mu)}
\end{equation}

\noindent with the flavor density matrices

\begin{equation}\label{flavorrho}
\hat{\rho}^{(\alpha)}= e^{-\beta(H_\alpha-\mu_{\alpha} q_\alpha)}
\end{equation}

Hence the initial distribution functions are given by

\begin{eqnarray}\label{occu}
\langle
\alpha_{\vk,\lambda}^{(\alpha)\dagger}\alpha_{\vk,\lambda}^{(\alpha)}
 \rangle &
= &  \mbox{Tr}
\hat{\rho}^{(\alpha)}\alpha_{\vk,\lambda}^{(\alpha)\dagger}\alpha_{\vk,\lambda}^{(\alpha)}
=
n^{(\alpha)}(k) \nonumber \\
\langle \beta_{\vk,
\lambda}^{(\alpha)\dagger}\beta_{\vk,\lambda}^{(\alpha)} \rangle &
= &  \mbox{Tr} \hat{\rho}^{(\alpha)}\beta_{\vk,
\lambda}^{(\alpha)\dagger}\beta_{\vk,\lambda}^{(\alpha)} =
\bar{n}^{(\alpha)}(k)
\end{eqnarray}

In the expressions above we have assumed that the distribution of
flavor neutrinos are spin independent, of course a spin dependence
of the distribution function can be incorporated in the
description.

Although we have stated the problem in terms of a gas flavor
neutrinos in thermal equilibrium  with Fermi-Dirac distributions,
this restriction can be relaxed to arbitrary non-equilibrium
single particle distributions consistent with Fermi-Dirac
statistics. Regardless of the initial distributions the ensuing
time evolution with the full Hamiltonian with mixing will be
\emph{out of equilibrium}.

\subsection{Cold degenerate case:} The case of a cold, degenerate
gas of neutrinos is described by the zero temperature limit but
fixed chemical potential of the density matrix (\ref{rhotot}) with
(\ref{flavorrho}). In this limit the individual flavor neutrino
gases form Fermi seas ``filled up'' to the Fermi momentum
$k_{F}^{(\alpha)}$. Consider the case of a positive chemical
potential corresponding to a degenerate gas of neutrinos without
antineutrinos at zero temperature, the degenerate ground state is
given by

\begin{equation}\label{FS}
|FS> = |FS>^{(e)}\otimes |FS>^{(\mu)}
\end{equation}

\noindent with

\begin{equation}
|FS>^{(\alpha)}=
\prod_{\vk}^{k_F^{(\alpha)}}\,\alpha_{\vk,\uparrow}^{(\alpha)\dagger}\alpha_{\vk,\downarrow}^{(\alpha)\dagger}|0>^{(\alpha)}
\end{equation}
\noindent with the flavor vacuum state $|0>^{(\alpha)}$
annihilated by the destruction operators
$\alpha_{\vk,\lambda}^{(\alpha)};\beta_{\vk,\lambda}^{(\alpha)}$.
The initial density matrix in this case is that of a pure state

\begin{equation}
\hat{\rho}= |FS><FS|
\end{equation}  the distribution function of flavor neutrinos
is given by

\begin{equation}\label{inidist}
n^{(\alpha)}(k) = \Theta(k_F^{(\alpha)}-k) \; , \;
\bar{n}^{(\alpha)}(k)=0
\end{equation}

\noindent and the chemical potential is
$\mu_{\alpha}=\omega_{\alpha}(k_F)$. The Fermi momentum is as
usual given by

\begin{equation}\label{KFs}
k_F^{(\alpha)} = \left(3\pi^2
\mathcal{N}^{(\alpha)}\right)^{1/3}\Rightarrow k_F^{(\alpha)}(eV)=
6.19 \left(\frac{\mathcal{N}^{(\alpha)}}{10^{15}\,\mbox{cm}^{-3}}
\right)^{1/3}
\end{equation}

\noindent with $\mathcal{N}^{(\alpha)}$ the neutrino density for
each flavor. Although the zero temperature limit is described by a
pure state, this state is a truly \emph{many body} state

An important \emph{many body} aspect of the situation under
consideration can be gleaned by studying how the creation and
annihilation operators of mass eigenstates act on the state
$|FS>$. Consider for example the action of the annihilation
operator $a^{(1)}_{\vk,\lambda}$ on the state, to understand this
question we must first obtain $a^{(1)}_{\vk,\lambda}$ in terms of
the creation and annihilation operators of flavor eigenstates.
From equation (\ref{psis}) and the relation between fields given
by (\ref{RotationMatrix}) we find

\begin{equation}
a^{(1)}_{\vk,\lambda} =
F^{(1),\dagger}_{\vk,\lambda}\left[C\nu_e(\vk)-S\nu_\mu(\vk)\right]
\end{equation}

\noindent and the expansion for the flavor fields given by
(\ref{diracflavor}) clearly indicates that  if $k<k_F^\mu<k_F^e$,
for example,  then $a^{(1)}_{\vk,\lambda}|FS>$ is a superposition
of states with an electron neutrino ``hole'', an electron
\emph{antineutrino}, a muon neutrino ``hole'' and a muon
\emph{antineutrino}. The antiparticle components of the wave
function $a^{(1)}_{\vk,\lambda}|FS>$ is a result of the
non-vanishing overlap between the positive energy spinors for mass
eigenstates and the \emph{negative} energy spinors for flavor
eigenstates\cite{blasone}.

\subsection{Time evolution}

Within the framework of free field theory of mixed neutrinos, the
time evolution is completely determined by the \emph{total
Hamiltonian} $H$ given by eqn. (\ref{totalham}).

In the Schroedinger picture the density matrix evolves in time
with the full Hamiltonian as follows

\begin{equation}\label{rhoevol}
\hat{\rho}(t) = e^{-iHt}\hat{\rho}(0)e^{iHt}
\end{equation}

Since the full Hamiltonian $H$ does not commute with $H_e,H_\mu$
because of the flavor mixing, the density matrix does not commute
with the Hamiltonian and therefore evolves in time. This is the
statement that the initial density matrix (\ref{rhotot}) describes
an ensemble \emph{out of equilibrium} when flavor neutrinos are
mixed.

Our goal is to obtain the time evolution of the distribution
functions for flavor neutrinos and antineutrinos, namely

\begin{equation}\label{noft}
n^{(\alpha)}(\vk,t) =   \mbox{Tr}
\hat{\rho}^{(\alpha)}(t)\,\alpha_{\vk,\lambda}^{(\alpha)\dagger}\alpha_{\vk,\lambda}^{(\alpha)}=
\mbox{Tr}
\hat{\rho}^{(\alpha)}(0)\,\alpha_{\vk,\lambda}^{(\alpha)\dagger}(t)\alpha_{\vk,\lambda}^{(\alpha)}(t)
\end{equation}

\noindent and similarly for the antineutrino distribution
function. The \emph{initial} distribution functions
$n^{(\alpha)}(\vk,0)=n^{(\alpha)}(\vk)$ (and similarly for
antineutrinos) given by equations (\ref{occu}) or (\ref{flavnumb})
for the case of an initial thermal distribution.

It is more convenient to describe the time evolution in the
Heisenberg picture wherein the density matrix does not depend on
time and the Heisenberg field operators carry the time dependence
as made explicit in eqn. (\ref{noft}).

The free fields associated with the mass eigenstates $\psi_i$
evolve in time with the usual time dependent phases multiplying
the creation and annihilation operators, namely
\begin{equation}\label{timeevolmass}
\psi_i(\vk,t) = e^{iHt} \psi_i(\vk,0) e^{-iHt}= \sum_{\lambda}
\left(
 a_{\vk, \lambda}^{(i)}\,e^{-iE_i(k)t}\,F_{\vk, \lambda}^{(i)}+
 b_{-\vk, \lambda}^{(i) \dagger}\,e^{iE_i(k)t}\,G_{-\vk, \lambda}^{(i)}
 \right)
\end{equation}

The time evolution of the fields associated with flavor
eigenstates, namely $\nu_{\alpha}$ is not so simple:

\begin{equation}\label{timeevolflav}
\nu_{\alpha}(\vk,t) = e^{iHt} \nu_{\alpha}(\vk,0) e^{-iHt}=
\sum_{\lambda}
 \left(
 \alpha_{\vk, \lambda}^{(\alpha)}(t)U^{(\alpha)}_{\vk,\lambda}+
 \beta_{-\vk , \lambda}^{(\alpha) \dagger}(t)V_{-\vk , \lambda}^{(\alpha)}
 \right)
\end{equation}

\noindent where the time dependent operators $\alpha_{\vk,
\lambda}^{(\alpha)}(t);\beta_{-\vk , \lambda}^{(\alpha)
\dagger}(t)$ can be obtained by writing the flavor fields in terms
of the mass eigenstate fields using eqn. (\ref{RotationMatrix})
and projecting out the components using the orthogonality property
given by eqn. (\ref{normaflavor}), leading for example to

\begin{eqnarray}\label{projec}
\alpha_{\vk, \lambda}^{(e)}(t) & = &
U^{(e)\dagger}_{\vk,\lambda}\left[C \psi_1(\vk,t)+S \psi_2(\vk,t)
\right] \nonumber \\
 \beta_{-\vk , \lambda}^{(e) \dagger}(t) & = & V^{(e)\dagger}_{-\vk,\lambda}\left[C \psi_1(\vk,t)+S \psi_2(\vk,t)
\right]
\end{eqnarray}

The expression (\ref{projec}) reveals several subtle aspects which
are highlighted by considering in detail for example the time
evolution of the  operator that creates electron neutrinos  (a
similar analysis holds for the muon neutrinos and their respective
antiparticles)

\begin{eqnarray}\label{ani}
\alpha^{(e)\dagger}_{\vk, \lambda}(t) & = & \sum_{\lambda'}
\left\{ \left( C \, a^{(1)\dagger}_{\vk, \lambda'}\,e^{iE_1(k)t}\,
F^{(1)\dagger}_{\vk, \lambda'}U^{(e)}_{\vk,\lambda}+ S \,
a^{(2)\dagger}_{\vk, \lambda'}\,e^{iE_2(k)t}\,
F^{(2)\dagger}_{\vk, \lambda'}U^{(e)}_{\vk,\lambda}\right) \right. +\nonumber \\
&  & \left. \left( C \,
 b_{-\vk, \lambda'}^{(1) }\,e^{-iE_1(k)t}\, G^{(1)\dagger}_{-\vk,
 \lambda'}U^{(e)}_{\vk,\lambda}+
 S  \, b_{-\vk, \lambda'}^{(2) }\,e^{-iE_2(k)t}\, G^{(2)\dagger}_{-\vk, \lambda'}U^{(e)}_{\vk,\lambda}
 \right)\right\}
\end{eqnarray}

It is a simple and straightforward exercise using the completeness
and orthogonality of the respective spinor wavefunctions, to show
that the creation and annihilation operators of flavor states
indeed fulfill the canonical anticommutation relations. A Fock
representation of flavor states is therefore consistent and
moreover \emph{needed} to describe a quantum statistical ensemble
of flavor neutrinos.

The first line in  the above expression shows that the
annihilation operator for electron corresponds to the expected
combination of creation operators for mass eigenstates multiplied
by the cosine and sine of the mixing angle, but also multiplied by
the overlap of the different spinor wavefunctions. Furthermore,
the electron creation operator also involves the
\emph{annihilation} of antiparticles associated with the mass
eigenstates, a feature recognized in ref.\cite{blasone}. There are
two important consequences of the \emph{exact} relation
(\ref{ani}):

\begin{itemize}
\item{ The amplitude for creating a mass eigenstate out of the
vacuum of mass eigenstates by an electron neutrino creation
operator is not only given by the cosine or sine (respectively) of
the mixing angle, but also by the overlap of the spinor wave
functions $F^{(i)\dagger}_{\vk, \lambda'}U^{(e)}_{\vk,\lambda}$.}
\item{The electron neutrino creation operator \emph{destroys}
antiparticle mass eigenstates. While this aspect is not relevant
when the electron neutrino creation operator acts on the
\emph{vacuum} of mass eigenstates, it becomes relevant in a medium
where both particles and antiparticles states are  populated.}
\end{itemize}

These aspects, which were also highlighted in
references\cite{blasone,fujii,ji} will be at the heart of the
subtle many body aspects of neutrino mixing which  contribute to
the time evolution of the distribution functions studied below.

The time dependent distribution functions are  obtained by taking
the trace with the initial density matrix

\begin{equation}\label{noft}
n^{(\alpha)}(\vk,t) =   \mbox{Tr}
\hat{\rho}^{(\alpha)}(0)\,\alpha_{\vk,\lambda}^{(\alpha)\dagger}(t)\,\alpha_{\vk,\lambda}^{(\alpha)}(t)
\end{equation}

 \noindent and similarly for the other distribution functions. One
 can use the expression (\ref{ani}) for the time evolution of
 the Heisenberg field operator (and the equivalent for the
 hermitian conjugate), however in order to compute the time
 evolved distribution function we would need to compute the
 expectation value of bilinears of the field operators $\psi_i$ in
 the \emph{flavor diagonal density matrix} $\hat{\rho}(0)$. To do
 this we would have to re-write the creation and annihilation
 operators $a_{\vk, \lambda}^{(i)}; b_{\vk,
 \lambda}^{(i)};\mbox{etc.}$ in the expression (\ref{ani}) back in terms of the creation and
 annihilation operators $\alpha_{\vk, \lambda}^{(\alpha)};\beta_{\vk,
 \lambda}^{(\alpha)};\mbox{etc.}$. This is obviously a rather
 cumbersome method. A more systematic manner to carry out this
 program is presented below.

Using the expressions (\ref{hamflavor},\ref{chargeflavor}) we find
the following identities

\begin{eqnarray}
&&\frac{1}{2}\langle \bar{\nu}_{\alpha}(\vk,t)\,\gamma^0\,
\nu_{\alpha}(\vk,t)\rangle   =   n^{(\alpha)}(\vk,t) -
\bar{n}^{(\alpha)}(\vk,t)+1 \label{diff} \\
&& \frac{1}{2 \omega_{\alpha}(k)}\langle
\bar{\nu}_{\alpha}(\vk,t)\,(\vec{\gamma}\cdot \vk + m_{\alpha})\,
\nu_{\alpha}(\vk,t)\rangle  =  n^{(\alpha)}(\vk,t) +
\bar{n}^{(\alpha)}(\vk,t)-1\label{plus}
\end{eqnarray}

Thus the computation of the distribution functions or combinations
of them requires to find general expressions of the form

\begin{equation}\label{expO}
<\bar{\nu}_{e}(\vk,t)\,\mathcal{O}\,\nu_{e}(\vk,t)>=\mathcal{O}_{fg}
<[\bar{\nu}_{e}(\vk,t)]_{f}[\nu_{e}(\vk,t)]_{g}>.
\end{equation}

\noindent where the Dirac indices $f,g$ are summed over and the
averages are in the flavor diagonal density matrix
(\ref{rhotot},\ref{flavorrho}).

 Since the time evolution of the fields $\psi_i$ is that of usual
 free Dirac field in terms of positive and negative frequency
 components,  we  write

\begin{equation}
\psi_{(i)}(\vk,t)=\left( \Lambda_{+}^{(i)}(\vk)\, e^{-iE_{i}t}+
\Lambda_{-}^{(i)}(\vk)\, e^{iE_{i}t} \right) \psi_{(i)}(\vk,0).
\label{psikt}
\end{equation}

Where we have introduced the  positive and negative frequency
projector operators $\Lambda_{+}(k)$ and $\Lambda_{-}(k)$
respectively which are given by

\begin{eqnarray}
\Lambda_{+}^{(i)}(\vk)& = &
\sum_{\lambda}F_{\vk,\lambda}^{(i)}F_{\vk,\lambda}^{(i)\dagger} =\left(\frac{\,\,\!\!\not\!k_{(i)}+M_{i}}{2 E_{i}}\right)\gamma^{0}, \label{Lambda+}\\
\Lambda_{-}^{(i)}(\vk)& = &
\sum_{\lambda}G_{-\vk,\lambda}^{(i)}G_{-\vk,\lambda}^{(i)\dagger} =\gamma^{0}\left(\frac{\,\,\!\!\not\!k_{(i)}-M_{i}}{2E_{i}}\right)\label{Lambda-}\\
 \!\!\not\!k_{(i)} & = & \gamma^{0}E_{i}(k)-\vec{\gamma}\cdot \vk
\end{eqnarray}

These projection operators have the following properties,

\begin{eqnarray}
&&\Lambda_{+}^{(i)\dagger}(\vk) = \Lambda_{+}^{(i)}(\vk)~~;~~ \Lambda_{-}^{(i)\dagger}(\vk) = \Lambda_{-}^{(i)}(\vk), \\
&&\Lambda_{+}^{(i)}(\vk) \, \Lambda_{-}^{(i)}(\vk) = 0 ~~;~~
\Lambda_{-}^{(i)}(\vk) \, \Lambda_{+}^{(i)}(\vk) = 0, \\
&&\Lambda_{+}^{(i)}(\vk)\,+\, \Lambda_{-}^{(i)}(\vk)=1.
\end{eqnarray}

We can now write the time evolution of the flavor fields in a
rather simple manner by using the relations between the fields
given by (\ref{RotationMatrix}) and the inverse relation which
allows to write $\psi_i(\vk,0)$ in (\ref{psikt}) back in terms of
$\nu_{\alpha}(\vk,0)$. We find

\begin{eqnarray}
\psi_1(\vk,t)  & =  & \gamma^0 F_1(\vk,t)[C \nu_{e}(\vk,0)-S
\nu_{\mu}(\vk,0)] \label{psi1t}\\
\bar{\psi}_1(\vk,t)  & =  & [C \bar{\nu}_{e}(\vk,0)-S
\bar{\nu}_{\mu}(\vk,0)] \tilde{F}_1(\vk,t)\gamma^0
\label{barpsi1t} \\
\psi_2(\vk,t)  & =  & \gamma^0 F_2(\vk,t)[C \nu_{\mu}(\vk,0)+S
\nu_{e}(\vk,0)] \label{psi2t}\\
\bar{\psi}_2(\vk,t)  & =  & [C \bar{\nu}_{\mu}(\vk,0)+S
\bar{\nu}_{e}(\vk,0)] \tilde{F}_2(\vk,t)\gamma^0 \label{barpsi2t}
\end{eqnarray}

Where we have introduced the following time evolution kernels

\begin{eqnarray}
F_{j}(\vk,t)&=&\gamma^{0}[\Lambda_{+}^{(j)}(\vk) e^{-iE_{j}(k)t}+
\Lambda_{-}^{(j)}(\vk) e^{iE_{j}(k)t}], \\
\tilde{F}_{j}(\vk,t) &=& F_{j}(\vk,-t)\gamma^{0}\; \; ; \; \;
j=1,2.
\end{eqnarray}

After straightforward algebra using the mixing transformation
(\ref{RotationMatrix}) and equations (\ref{psi1t}-\ref{barpsi2t})
we find the following result for the time evolution of the flavor
fields

\begin{eqnarray}
\nu_{e}(k,t)&=&T_{ee}(\vk,t)\nu_{e}(\vk,0)+T_{e\mu}(\vk,t)\nu_{\mu}(\vk,0), \label{nuekt} \\
\bar{\nu}_{e}(k,t)&=&\bar{\nu}_{e}(\vk,0)\tilde{T}_{ee}(\vk,t)+\bar{\nu}_{\mu}(\vk,0)\tilde{T}_{e\mu}(\vk,t),
\label{nubarekt}\\
\nu_{\mu}(\vk,t)&=& T_{\mu\mu}(\vk,t)\nu_{\mu}(\vk,0)+T_{\mu e}(\vk,t)\nu_{e}(\vk,0), \label{numukt}  \\
\bar{\nu}_{\mu}(\vk,t)&=&
\bar{\nu}_{\mu}(\vk,0)\tilde{T}_{\mu\mu}(\vk,t)+\bar{\nu}_{e}(\vk,0)\tilde{T}_{\mu
e}(\vk,t), \label{numubarkt}
\end{eqnarray}

where the time evolution operators are given by

\begin{eqnarray}
T_{ee}(\vk,t)&=& \gamma^0\left[C^2 F_1(\vk,t) +
S^2 F_2(\vk,t)\right], \\
T_{\mu\mu}(\vk,t)&=& \gamma^0\left[C^2 F_2(\vk,t) + S^2
F_1(\vk,t)\right], \\
T_{e\mu}(\vk,t)&=& T_{\mu e}=CS \gamma^0
\left[F_2(\vk,t)-F_1(\vk,t)\right] \\
\tilde{T}_{\alpha \beta}(\vk,t)&=&\gamma^{0}T_{\alpha
\beta}(\vk,-t)\gamma^{0},
\end{eqnarray}

Furthermore since the initial density matrix is flavor diagonal,
we find the following expectation values

\begin{eqnarray}
<[\bar{\nu}_{e}(\vk,t)]_{f}[\nu_{e}(\vk,t)]_{g}>&=&
<[\bar{\nu}_{e}(\vk,0)]_{r}[\nu_{e}(\vk,0)]_{s}>[\tilde{T}_{ee}(\vk,t)]_{rf}[T_{ee}(\vk,t)]_{gs}
\nonumber \\
&&+<[\bar{\nu}_{\mu}(\vk,0)]_{r}[\nu_{\mu}(\vk,0)]_{s}>[\tilde{T}_{e\mu}(\vk,t)]_{rf}[T_{e\mu}(\vk,t)]_{gs},
\end{eqnarray}

\noindent and similarly for the muon neutrino fields, where
$<\cdots >$ stands for the trace with the initial density matrix.

A noteworthy feature of the above \emph{exact} expressions is that
the time evolution of the flavor neutrino fields \emph{mix
positive and negative frequency} components of the mass
eigenstates. Namely a flavor neutrino state is a linear
combination of particles and \emph{antiparticles} of mass
eigenstates. Thus a wave packet of flavor neutrinos will
necessarily mix positive and negative frequencies of mass
eigenstates. This mixing between particles and antiparticles is a
consequence of the fact that a flavor eigenstate is a squeezed
state of mass eigenstates and viceversa\cite{blasone}.

A simple calculation yields the following expectation values in
the initial density matrix

\begin{eqnarray}
<[\bar{\nu}_{\alpha}(\vk,0)]_{r}[\nu_{\alpha}(\vk,0)]_{s}> &=&
\left[\sum_{\lambda}<\alpha_{\vk,\lambda}^{(\alpha)\dagger}\alpha_{\vk,\lambda}^{(\alpha)}>
[\bar{U}_{\vk,\lambda}^{(\alpha)}]_{r}[U_{\vk,\lambda}^{(\alpha)}]_{s}
+
\sum_{\lambda}<\beta_{-\vk,\lambda}^{(\alpha)\dagger}\beta_{-\vk,\lambda}^{(\alpha)}>
[\bar{V}_{-\vk,\lambda}^{(\alpha)}]_{r}[V_{-\vk,\lambda}^{(\alpha)}]_{s} \right]\\
&=& n^{(\alpha)}(k)
\left(\frac{\!\!\not\!k_{\alpha}+m_{\alpha}}{2\omega_{\alpha}(k)}\right)_{sr}
+(1-\bar{n}^{(\alpha)}(k))\left[\gamma^{0}\frac{\,\,\!\!\not\!k_{\alpha}-m_{\alpha}}{2\omega_{\alpha}(k)}\gamma^{0}\right]_{sr}
\equiv [N_{\alpha}(\vk)]_{sr}\\
\!\!\not\!k_{\alpha}& = &
\gamma^{0}\omega^{\alpha}(k)-\vec{\gamma}\cdot \vk
\end{eqnarray}

\noindent where $n^{\alpha}(k);\bar{n}^{\alpha}(k)$ are given by
the expressions (\ref{flavnumb}) and there are no flavor
off-diagonal matrix elements at $t=0$ because the initial density
matrix  is flavor diagonal.

Combining all the above results, we find the final compact form
for the time dependent expectation values in eqn. (\ref{expO}),
namely

\begin{equation}
<\bar{\nu}_{e}(\vk,t)\,\mathcal{O}\,\nu_{e}(\vk,t)>  =
Tr\left[N_{e}(\vk)\tilde{T}_{ee}(\vk,t)\,\mathcal{O}\,T_{ee}(\vk,t)\right]
+Tr\left[
N_{\mu}(\vk)\tilde{T}_{e\mu}(\vk,t)\,\mathcal{O}\,T_{e\mu}(\vk,t)\right]
\end{equation}

\subsubsection{Exact time evolution of distribution functions}

The exact time evolution (in free field theory) of flavor
neutrinos is given by

\begin{equation}\label{nel}
n^{(e)}(k,t)\equiv I^{(e)}(k,t)+J^{(e)}(k,t),
\end{equation}

where $I^{(e)}(k,t)$ and $J^{(e)}(k,t)$ are given by

\begin{eqnarray}
I^{(e)}(k,t)&=& \frac{1}{4\omega_{e}(k)}Tr\left[
N_{e}(k)\tilde{T}_{ee}(\vk,t)\gamma^{0}(\,\,\!\!\not\!k_{e}+m_{e})\gamma^{0}T_{ee}(\vk,t)\right], \\
J^{(e)}(k,t)&=& \frac{1}{4\omega_{e}(k)}Tr\left[
N_{\mu}(\vk)\tilde{T}_{e\mu}(\vk,t)\gamma^{0}(\,\,\!\!\not\!k_{e}+m_{e})\gamma^{0}T_{e\mu}(\vk,t)\right].
\end{eqnarray}

\begin{equation}\label{nbarel}
\bar{n}^{(e)}(k,t)=1-\bar{I}^{(e)}(k,t)-\bar{J}^{(e)}(k,t),
\end{equation}

where $\bar{I}^{(e)}(k,t)$ and $\bar{J}^{(e)}(k,t)$ are given by

\begin{eqnarray}
\bar{I}^{(e)}(k,t)&=& \frac{1}{4\omega_{e}(k)}Tr\left[
N_{e}(\vk)\tilde{T}_{ee}(\vk,t)(\,\,\!\!\not\!k_{e}-m_{e})T_{ee}(\vk,t)\right], \\
\bar{J}^{(e)}(k,t)&=& \frac{1}{4\omega_{e}(k)}Tr\left[
N_{\mu}(\vk)\tilde{T}_{e\mu}(\vk,t)(\,\,\!\!\not\!k_{e}-m_{e})T_{e\mu}(\vk,t)\right].
\end{eqnarray}

For the muon neutrinos and antineutrinos

\begin{equation}
n^{(\mu)}(k,t)=I^{(\mu)}(k,t)+J^{(\mu)}(k,t)
\end{equation}

where $I^{(\mu)}(k,t)$ and $J^{(\mu)}(k,t)$ are given by

\begin{eqnarray}
I^{(\mu)}(k,t)&=& \frac{1}{4\omega_{\mu}(k)}Tr\left[
N_{\mu}(\vk)\tilde{T}_{\mu\mu}(\vk,t)\gamma^{0}(\,\,\!\!\not\!k_{\mu}+m_{\mu})\gamma^{0}T_{\mu\mu}(\vk,t)\right], \\
J^{(\mu)}(k,t)&=& \frac{1}{4\omega_{\mu}(k)}Tr\left[
N_{e}(\vk)\tilde{T}_{\mu
e}(\vk,t)\gamma^{0}(\,\,\!\!\not\!k_{\mu}+m_{\mu})\gamma^{0}T_{\mu
e}(\vk,t)\right].
\end{eqnarray} \\

\begin{equation}\label{nbarel}
\bar{n}^{(\mu)}(k,t)=1-\bar{I}^{(\mu)}(k,t)-\bar{J}^{(\mu)}(k,t),
\end{equation}

where $\bar{I}^{(\mu)}(k,t)$ and $\bar{J}^{(\mu)}(k,t)$ are given
by

\begin{eqnarray}
\bar{I}^{(\mu)}(k,t)&=& \frac{1}{4\omega_{\mu}(k)}Tr\left[
N_{\mu}(\vk)\tilde{T}_{\mu \mu}(\vk,t)(\,\,\!\!\not\!k_{\mu}-m_{\mu})T_{\mu \mu}(\vk,t)\right], \\
\bar{J}^{(\mu)}(k,t)&=& \frac{1}{4\omega_{\mu}(k)}Tr\left[
N_{e}(\vk)\tilde{T}_{\mu
e}(\vk,t)(\,\,\!\!\not\!k_{\mu}-m_{\mu})T_{\mu e}(\vk,t)\right].
\label{muqs}
\end{eqnarray}

The calculation of the traces is simplified by the observation
that all of the different terms that enter in the trace, such as
$N_\alpha(\vk); \tilde{T}_{\alpha, \alpha'}(\vk,t)\gamma^0;
\gamma^0 {T}_{\alpha, \alpha'}(\vk,t)$ can be written in the form

\begin{equation}
\gamma^0 A_0(\vk,t)-\vec{\gamma}\cdot \vec{A}(\vk,t)+B(\vk,t)
\equiv \; \!\not\!\!A(\vk,t) + B(\vk,t)
\end{equation}

\noindent where the coefficient functions
$A_0(\vk,t);\vec{A}(\vk,t); B(\vk,t)$ can be read off each
individual term. Thus the traces in the terms above can be
calculated by using the standard formulae for the traces of two
and four Dirac matrices.

\subsection{Fast and slow time scales}\label{timescales}

While the exact compact expressions above describe  the full time
evolution and provide a set of closed form expressions, they hide
the fact that there two  \emph{widely different} time scales.
These different time scales can be revealed by unravelling the
different contributions to the distribution functions as follows.
Consider the expectation value on the right hand side of eqn.
(\ref{expO}) for the case of the electron neutrino

\begin{eqnarray}\label{split}
<[\bar{\nu}_{e}(\vk,t)]_{f}[\nu_{e}(\vk,t)]_{g}> & = & C^2
<[\bar{\psi}_{1}(\vk,t)]_{f}[\psi_{1}(\vk,t)]_{g}>+ S^2
<[\bar{\psi}_{2}(\vk,t)]_{f}[\psi_{2}(\vk,t)]_{g}> \nonumber \\
& +& C S
<[\bar{\psi}_{1}(\vk,t)]_{f}[\psi_{2}(\vk,t)]_{g}+[\bar{\psi}_{2}(\vk,t)]_{f}[\psi_{1}(\vk,t)]_{g}>
\end{eqnarray}

\noindent the case of the muon neutrino can be obtained from the
expression above by replacing $S\rightarrow C; C\rightarrow -S$.

By writing each one of the fields $\psi_i$ in terms of the
positive and negative frequency contributions which evolve in time
with the phases $e^{\mp iE_i(k)t}$ respectively, it is clear that
in the products $\bar{\psi}_i(\vk,t)\, \psi_i(\vk,t)$ there is a
contribution that does not depend on time and terms that oscillate
in time with the phases  $e^{\mp 2 i E_i(k)t}$. These oscillatory
terms arise from the interference between particles and
antiparticles akin to \emph{zitterbewegung} in principle do not
vanish when the density matrix is diagonal in the flavor basis. In
the general expectation values in eqn (\ref{expO}) these
oscillatory terms will multiply the matrix elements of the form
$\bar{F}^{(i)}_{\vk,\lambda}\mathcal{O}G^{(i)}_{-\vk,\lambda}$,
thus if these matrix elements do not vanish, these oscillatory
terms are present. In the second line in eqn. (\ref{split}) a
similar argument shows that there are two types of oscillatory
terms, $e^{\mp i (E_1(k)+E_2(k))t} $ and $e^{\mp i
(E_1(k)-E_2(k))t} $. The former arise from the interference
between the particle and antiparticle states of different masses,
while the latter from interference between particle states of
different masses (or antiparticle).

The combined analysis from solar neutrinos and
KamLAND\cite{kamland} suggest that for two flavor mixing
$M^2_1-M^2_2=\Delta M^2 \sim 7 \times 10^{-5} (eV)^2$ and
cosmological constraints from WMAP\cite{WMAP} suggest that the
average mass of neutrinos is $\bar{M} \lesssim 0.23 eV$. Therefore
even in the non-relativistic limit with $k\ll M_i$ the ratio
$|E_1(k)-E_2(k)|/(E_1(k)+E_2(k)) < 10^{-4}$ and certainly much
smaller in the relativistic limit $k \gg M_i$. Hence  because of
the near degeneracy, or in the relativistic limit for any value of
the masses, there are two widely different time scales of
evolution for the flavor distribution functions. The longest one
corresponding to the interference between particle states (or
antiparticle states) of different masses while the shortest one
corresponds to the interference between particle and antiparticle
states of equal or different masses. This point will be revisited
below.

The evolution of the flavor (lepton) \emph{asymmetry} highlights
these time scales clearly and is studied below.

\section{Degenerate gas of neutrinos: evolution of flavor asymmetry}

The results obtained above are general and valid for any
temperature and chemical potential (density). In this section we
focus on understanding the time evolution of the flavor asymmetry
$n^{(\alpha)}(k,t)-\bar{n}^{(\alpha)}(k,t)$ in the case of a cold,
degenerate gas of flavor neutrinos. From equations (\ref{diff})
and (\ref{split}) we find

\begin{eqnarray}
n^{(e)}(\vk,t)-\bar{n}^{(e)}(\vk,t) & = & \frac{C^2}{2}
<\psi_1^{\dagger}(\vk,t)\psi_1(\vk,t)>+\frac{S^2}{2}
<\psi_2^{\dagger}(\vk,t)\psi_2(\vk,t)>-1\nonumber \\& + &
\frac{CS}{2}<\psi_1^{\dagger}(\vk,t)\psi_2(\vk,t)+\psi_2^{\dagger}(\vk,t)\psi_1(\vk,t)>
\label{eleasy}\\
n^{(\mu)}(\vk,t)-\bar{n}^{(\mu)}(\vk,t) & = & \frac{S^2}{2}
<\psi_1^{\dagger}(\vk,t)\psi_1(\vk,t)>+\frac{C^2}{2}
<\psi_2^{\dagger}(\vk,t)\psi_2(\vk,t)>-1\nonumber\\ & - &
\frac{CS}{2}<\psi_1^{\dagger}(\vk,t)\psi_2(\vk,t)+\psi_2^{\dagger}(\vk,t)\psi_1(\vk,t)>
\label{muasy}
\end{eqnarray}

The first line of the expressions above is time independent
because the overlap between positive and negative frequency
components vanishes, and the time dependence arises solely from
the interference between different mass eigenstates. The time
dependent terms (second lines in the above expressions) are
opposite for the two flavors realizing the fact that the total
charge of mass eigenstates equals that of flavor eigenstates  and
is time independent (see eqn. (\ref{totalcharge})).

Furthermore the expectation values
$<\psi_i^{\dagger}(\vk,t)\,\psi_i(\vk,t)>$ (no sum on $i$) are
time independent (in the case of free field theory under
consideration) since the interference term between positive and
negative frequency spinors vanishes. The time dependence is
completely encoded in the contribution that mixes the mass
eigenstates.

Therefore the time dependence of the flavor asymmetry is
completely determined by the quantity

\begin{equation}
\chi(\vk, t) \equiv \frac{CS}{2} <\left(
\psi_{1}^{\dagger}(\vk,t)\psi_{2}(\vk,t)+
\psi_{2}^{\dagger}(\vk,t)\psi_{1}(\vk,t)\right)>.
\end{equation}

Using equations (\ref{psi1t})- (\ref{barpsi2t}), it follows that

\begin{eqnarray}
\chi(\vk,t) = \frac{C^{2}S^{2}}{2}\,
Tr\left[\,[N_{e}(\vk)-N_{\mu}(\vk)][\tilde{F}_{1}(\vk,t)F_{2}(\vk,t)+
\tilde{F}_{2}(\vk,t)F_{1}(\vk,t)]\,\right],
\end{eqnarray}

The computation of the traces is simplified by writing

\begin{eqnarray}
F_{j}(k,t)& = & \,\,\!\!\not\!P_{j}(t)+M_{j}(t) \label{Fjoft}\\
P_{j}^{0}(t)&=& \cos(E_{j}(k)t), \\
\vec{P}_{j}(t) &=& \frac{i\vec{k}}{E_{j}(k)}\sin(E_{j}(k)t), \\
M_{j}(t)&=&-\frac{iM_{j}}{E_{j}(k)}\sin(E_{j}(k)t).
\end{eqnarray}

\noindent and  similarly we write

\begin{eqnarray}
N_{\alpha}(k)& = &
\,\,\!\!\not\!Q_{\alpha}+\widetilde{M}_{\alpha}\label{Nalfa}\\
\!\!\not\!Q_{\alpha}&=&\gamma^{0}Q_{\alpha}^{0}-\vec{\gamma}\cdot \vec{Q}_{\alpha}, \\
Q_{\alpha}^{0}&=&\frac{1}{2}\left[n^{(\alpha)}(k)+1-\bar{n}^{(\alpha)}(k)\right], \\
\vec{Q}_{\alpha}&=&\frac{\vec{k}}{2\omega_{\alpha}(k)}\left[n^{(\alpha)}(k)-1+\bar{n}^{(\alpha)}(k)\right], \\
\widetilde{M}_{\alpha}&=&\frac{m_{\alpha
}}{2\omega_{\alpha}(k)}\left[n^{(\alpha)}(k)-1+\bar{n}^{(\alpha)}(k)\right].
\end{eqnarray}

For further convenience, we define

\begin{equation}
\Delta\!\!\not\!Q = \!\!\not\!Q_{e}-\!\!\not\!Q_{\mu},~~;~~ \Delta
\widetilde{M} = \widetilde{M}_{e}-\widetilde{M}_{\mu},
\end{equation}

\noindent in terms of which we obtain

\begin{eqnarray}
\chi(\vk,t)&=& \frac{C^{2}S^{2}}{2} \,\,
Tr\left[\,(\Delta\!\!\not\!Q+\Delta\widetilde{M})
(\,\,\!\!\not\!P_{1}(-t)+M_{1}(-t))\gamma^{0}
(\,\,\!\!\not\!P_{2}(t)+M_{2}(t)) \right. \nonumber \\
&&\left. +(\Delta\!\!\not\!Q+\Delta\widetilde{M})
(\,\,\!\!\not\!P_{2}(-t)+M_{2}(-t))\gamma^{0}
(\,\,\!\!\not\!P_{1}(t)+M_{1}(t))\,\right].
\end{eqnarray}

After some lengthy but straigthforward algebra we find

\begin{eqnarray}\label{chioft}
\chi(\vk, t) & = &\chi(\vk,0)-2C^{2}S^{2}
\,\,\left[\left(n^{(e)}(k)-\bar{n}^{(e)}(k)\right)-\left(n^{(\mu)}(k)-\bar{n}^{(\mu)}(k)\right)\right]\,\Bigg[
\left(1-\frac{k^{2}+M_{1}M_{2}}{E_{1}(k)E_{2}(k)}\right)\sin^{2}\Bigg(\frac{E_{1}(k)+E_{2}(k)}{2}~t\Bigg)+
\nonumber \\ & &
\left(1+\frac{k^{2}+M_{1}M_{2}}{E_{1}(k)E_{2}(k)}\right)\sin^{2}\Bigg(\frac{E_{1}(k)-E_{2}(k)}{2}~t\Bigg)
\Bigg],
\end{eqnarray}

where $\chi(\vk,0)$ is given by

\begin{equation}
\chi(\vk,0)=2C^{2}S^{2}
\,\,\left[\left(n^{(e)}(k)-\bar{n}^{(e)}(k)\right)-\left(n^{(\mu)}(k)-\bar{n}^{(\mu)}(k)\right)\right].
\end{equation}

The expression (\ref{chioft}) for the time dependence of the
flavor asymmetry  clearly shows that neutrino mixing results in a
time evolution of the flavor asymmetry \emph{unless} the flavor
asymmetry  for both flavors is the same. This is obviously a
consequence of Pauli blocking: if the neutrino states are occupied
up to the same momentum electron neutrinos cannot transform into
an (occupied) muon neutrino state and viceversa.

In the case of a cold, degenerate gas of flavor neutrinos (we
assume here both chemical potentials to be positive) is given by

\begin{equation} \label{dege}
n^{(\alpha)}(k)   \rightarrow   \Theta(k^{(\alpha)}_F-k) \; ; \;
\bar{n}^{(\alpha)}(k)   \rightarrow  0
\end{equation}

If the chemical potential is different for the different flavors,
the expression above shows that each wavevector mode will evolve
with a different frequency and as a consequence of free field
evolution there is no mode mixing. The important question is what
is the time evolution of the \emph{total charge} which is the
integral of the flavor asymmetry over all momenta. This time
evolution will be a result of the \emph{dephasing} through the
oscillations between different modes that are not Pauli blocked.

We now proceed to study analytically and  numerically  the time
evolution of the  flavor charge densities $q_{\alpha}/V$ with
$q_{\alpha}$ given by eqn. (\ref{chargeflavor}) and $V$ the
volume. We begin by defining

\begin{equation}\label{defmass}
\bar{M}\equiv \frac{M_{1}+M_{2}}{2} ~~;~~ \Delta M^{2}\equiv
M_{1}^{2}-M_{2}^{2},
\end{equation}

\noindent so that $M_{1}$ and $M_{2}$ can be written in terms of
$\bar{M}$ and $\Delta M^{2}$ as

\begin{equation}\label{massmass}
M_{1}=\bar{M}\left(1+\frac{\Delta M^{2}}{4\bar{M}^{2}}\right)~~;~~
M_{2}=\bar{M}\left(1-\frac{\Delta M^{2}}{4\bar{M}^{2}}\right).
\end{equation}

We take the following as representative values for the two flavor
case\cite{smirnov,giunti} $\bar{M}\simeq 0.25~eV$ and $\Delta
M^{2}\simeq 7\times 10^{-5}(eV)^{2}$. In what follows we assume
that $k_{F}^{e}>k_{F}^{\mu}$ and introduce dimensionless variables
by taking $k_{F}^{e}$ as the common scale, the opposite limit for
the Fermi momenta can be obtained simply from the results below.
Hence we define

\begin{eqnarray}\label{dimvars}
q&=&\frac{k}{k_{F}^{e}}\; ; \;
q_{r}=\frac{k_{F}^{\mu}}{k_{F}^{e}}\; ; \;
\tau=k_{F}^{e} t, \\
\bar{m} & = & \frac{\bar{M}}{k_{F}^{e}}~~;~~ \delta m^2= m^2_1-m^2_2 = \frac{M^2_1-M^2_2}{(k^{e}_F)^2} \\
m_{1}&=&\bar{m}\left(1+\frac{\Delta M^{2}}{4\bar{M}^{2}}\right)\;
; \; m_{2}=\bar{m}\left(1-\frac{\Delta M^{2}}{4\bar{M}^{2}}\right)
; \\ \varepsilon_{1}&=&\sqrt{q^{2}+m_{1}^{2}}\; ; \;
\varepsilon_{2}=\sqrt{q^{2}+m_{2}^{2}}.
\end{eqnarray}

Hence, in terms of
$\mathcal{N}^{(\alpha)}=(k_{F}^{\alpha})^{3}/3\pi^{2}$ (see eqn.
(\ref{KFs})), we find that the time evolution of the flavor charge
densities are given by

\begin{eqnarray}
\frac{q_e(t)}{V} & = & \mathcal{N}^{(e)} -6 C^2 S^2
\mathcal{N}^{(e)} \Big(
I_f(\tau)+I_s(\tau)\Big) \label{qeleoft} \\
\frac{q_\mu(t)}{V} & = & \mathcal{N}^{(\mu)}+6 C^2 S^2
\mathcal{N}^{(e)}\Big( I_f(\tau)+I_s(\tau)\Big) \label{qmuoft}
\end{eqnarray}

\noindent where

\begin{eqnarray}
I_f(\tau) &=&\int_{q_{r}}^{1}dq \, q^{2}
\left(1-\frac{q^{2}+m_{1}m_{2}}{\varepsilon_{1}\varepsilon_{2}}\right)
\sin^{2}\Bigg[\frac{\varepsilon_{1}+\varepsilon_{2}}{2}\,\tau\Bigg]\label{Ifast}\\
I_s(\tau) & = & \int_{q_{r}}^{1}dq \, q^{2}
\left(1+\frac{q^{2}+m_{1}m_{2}}{\varepsilon_{1}\varepsilon_{2}}\right)
\sin^{2}\Bigg[\frac{\varepsilon_{1}-\varepsilon_{2}}{2}\,\tau\Bigg]
\label{Islow}
\end{eqnarray}

We have separated the contributions from the fast ($I_f(\tau)$)
and slow ($I_s(\tau)$) time scales as discussed in section
(\ref{timescales}) above. In particular, as discussed above the
term that oscillates with the \emph{sum}
$\varepsilon_{1}+\varepsilon_{2}$ is a consequence of the overlap
between particles and antiparticles. The pre-factors that multiply
the sine functions in equations (\ref{Ifast},\ref{Islow})  arise
from the overlap between \emph{particle-antiparticle} spinors in
(\ref{Ifast}) and particle-particle, anti-particle-anti-particle
spinors in (\ref{Islow}).    The overlap between particle and
antiparticle spinors is non-vanishing for different masses.
Similar contributions from the overlap between particle and
antiparticle states of different masses have been found in the
studies of refs.\cite{blasone,fujii,ji}.

Since the mass eigenstates are almost degenerate or alternatively
for any values of the masses in the relativistic limit we find

\begin{eqnarray}\label{approx}
\frac{q^{2}+m_{1}m_{2}}{\varepsilon_{1}\varepsilon_{2}} & = &
1-\frac{\bar{m}^2 q^2 }{\bar{\varepsilon}^4} \left(\frac{\Delta
M^{2}}{4\bar{M}^{2}}\right)^2 + \mathcal{O}\left(
\left(\frac{\Delta
M^{2}}{4\bar{M}^{2}}\right)^4 \right)\nonumber\\
\bar{\varepsilon} & = & \sqrt{q^2+\bar{m}^2}
\end{eqnarray}

\noindent with $\frac{\Delta M^{2}}{4\bar{M}^{2}} \sim 3 \times
10^{-4}$. Therefore  the coefficient that results from the overlap
between the particle and antiparticle spinors of different mass is
given by

\begin{equation}
1-\frac{q^{2}+m_{1}m_{2}}{\varepsilon_{1}\varepsilon_{2}} =
\mathcal{O}\left( \frac{\Delta
M^{2}}{4\bar{M}^{2}}\frac{\bar{M}}{E(k)} \right)^2
\label{partianti}
\end{equation}

\noindent and the coefficient that results from the overlap
between particle-particle or anti-particle-anti-particle of
\emph{different} masses is

\begin{equation}
1+\frac{q^{2}+m_{1}m_{2}}{\varepsilon_{1}\varepsilon_{2}} = 2+
\mathcal{O}\left( \frac{\Delta
M^{2}}{4\bar{M}^{2}}\frac{\bar{M}}{E(k)} \right)^2
\label{partparti}
\end{equation}

\noindent where $E(k)$ is an energy scale.

 Therefore the coefficient of the oscillatory term in $I_f(\tau)$ is a factor at least of order $\left(\frac{\Delta
M^{2}}{4\bar{M}^{2}}\right)^2 \sim 10^{-7}$ smaller than that of
$I_s(\tau)$. Furthermore it is clear that the interference terms
between particle and antiparticle average out  on a time scale
$t_f \lesssim 1/{\bar{M}}$ whereas the particle-particle
contributions evolve  on a much slower time scale $t_s \sim
\bar{M}/\Delta M^2\gg t_f$.

However, despite the fact that the coefficients of the oscillatory
terms in $I_s(\tau)$ and $I_f(\tau)$ differ by several orders of
magnitude, the fact that the time evolution of $I_s(\tau)$ is much
slower allows for a time scale within which both contributions are
\emph{comparable}. This can be gleaned from the following
argument.

The integrals for $I_s(\tau)$ and $I_f(\tau)$ are dominated by the
region $q \sim 1$. Consider an intermediate time scale so that the
argument of the oscillatory function in $I_f(\tau)$ is of order
one, but the argument of the oscillatory function in $I_s(\tau)$
is $\ll 1$. The contribution to the integral in $I_f(\tau)$ is of
order $\bar{m}^2  \left(\frac{\Delta m^{2}}{4\bar{m}^{2}}\right)^2
$ while the contribution to the integral $I_s(\tau)$ is of order $
2 (\delta m^2 \tau^2) $. Therefore,  it is  clear that even when
the prefactor of its oscillatory term is small, the integrand of
$I_f(\tau)$ will be \emph{larger than} that of $I_s(\tau)$ in the
time domain during which

\begin{equation}\label{ineq}
\bar{m}^2 \left( \frac{\delta m^2}{4\bar{m}^2} \right)^2 > (\delta
m^2 \tau)^2 \Longrightarrow \tau \lesssim 1/\bar{m}
\end{equation}

In the opposite limit, for $\tau >> 1/\bar{m}$ the dynamics is
completely dominated by $I_s(\tau)$.

Fig. (\ref{fig:shortime}) below displays the early time evolution
of $I_s(\tau)$ and $I_f(\tau)$ for $0 \leq \tau \lesssim
1/\bar{m}$. It is clear from this figure that $I_f(\tau)$ averages
out to its asymptotic value on a short time scale $\tau \sim 1$
($t \sim 1/k_F$) and that $I_s(\tau)$ begins to dominate the
dynamics on time scales $\tau \gtrsim 1/\bar{m}$ as discussed
above. In the case of Fig.(\ref{fig:shortime}), with
$k^e_F>>\bar{M}$ the time scale of averaging is $t \sim 1/k^e_F$,
but for $k_F << \bar{M}$ it would be of order $1/\bar{M}$.

\begin{figure}[ht!]
\begin{center}
\includegraphics[height= 8cm,width=8cm,keepaspectratio=true]{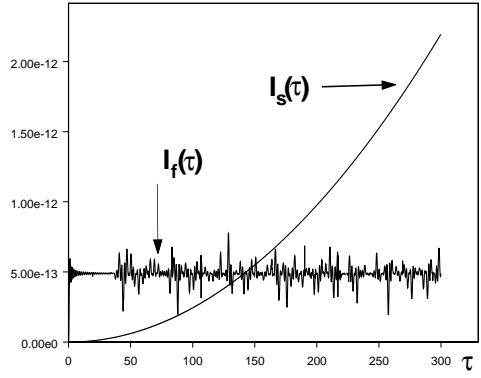}
\caption{$I_s(\tau)$ and $I_s(\tau)$ for $k^e_F=100 ~eV\; ; \;
k^\mu_F=0\; ; \; \bar{M}=0.25 ~ eV\; ; \; \Delta M^{2}\simeq
7\times 10^{-5}(eV)^{2}$ vs. $ \tau$. For these values $1/\bar{m}
= 400$.} \label{fig:shortime}
\end{center}
\end{figure}

In terms of dimensionful  quantities the inequality obtained in
eqn. (\ref{ineq}) above translates into $t < 1/\bar{M}$. With the
current estimate $\bar{M} \sim 0.25~\mbox{eV}$ the analysis above
suggests that the particle-antiparticle interference is
dynamically relevant during time scales $t \lesssim 10^{-15}~s$
although this time scale is comparable to the expansion time scale
at the time of the electroweak phase transition, it is far shorter
than the time scales relevant  either for primordial
nucleosynthesis or for dynamical processes during the collapse of
supernovae or neutron star cooling.

While the behavior of $I_s(\tau)$ and $I_f(\tau)$ as a function of
$\tau$ must in general be studied numerically, the long time limit
can be extracted analytically.

The asymptotic long time behavior of $I_s(\tau)$ and $I_f(\tau)$
is determined by the end points of their integrands, in particular
for momenta near the Fermi surface. Two relevant cases yield the
following results

\begin{itemize}
\item{{ \bf{Relativistic Limit:} $\mbox{max}(k^e_F,k^\mu_F) \gg
M_1,M_2$}
 \begin{eqnarray}
 I_s(\tau) & = &  \frac{1}{2}\int_{q_{r}}^{1}dq \, q^{2}
\left(1+\frac{q^{2}+m_{1}m_{2}}{\varepsilon_{1}\varepsilon_{2}}\right)+
\frac{2}{\delta m^2\tau}\left\{\sin\Big(\frac{\delta
m^2}{2}\tau\Big)-q^4_r \sin\Big(\frac{\delta m^2\tau}{2q_r}\Big)
\right\}+
\mathcal{O}\left(\frac{1}{(\delta m^2 \tau)^2}\right) \label{Islowrel}\\
I_f(\tau) & = &  \frac{1}{2}\int_{q_{r}}^{1}dq \, q^{2}
\left(1-\frac{q^{2}+m_{1}m_{2}}{\varepsilon_{1}\varepsilon_{2}}\right)-
\frac{1}{8\tau}(m_1-m_2)^2\left[\sin(2\tau)-\sin{2q_r\tau}\right]+\mathcal{O}\left(\frac{1}{\tau^2}\right)
\label{Ifastrel}
\end{eqnarray}
 \noindent where $\delta m^2$ is defined by equation
(\ref{dimvars}) along with the other dimensionless variables. }

\item{{\bf{Non-Relativistic limit:} $k^e_F, k^\mu_F \ll M_1,M_2$}
\begin{eqnarray}
 I_s(\tau) & = &  \frac{1}{2}\int_{q_{r}}^{1}dq \, q^{2}
\left(1+\frac{q^{2}+m_{1}m_{2}}{\varepsilon_{1}\varepsilon_{2}}\right)+
\frac{m_1m_2}{(m_1-m_2)~\tau}\left\{\sin\Big[(m_1-m_2)(1-\frac{1}{2m_1m_2})\tau\Big]-
\right.
\nonumber \\
&  & \left. q_r
\sin\Big[(m_1-m_2)(1-\frac{q^2_r}{2m_1m_2})\tau\Big]\right\}+
\mathcal{O}\left(\frac{1}{\tau^{\frac{3}{2}}}\right)
~~;~~\mbox{for}~~
\frac{(m_1-m_2)\tau}{m_1m_2}\gg 1 \nonumber \\
I_s(\tau) & = &  \frac{1}{2}\int_{q_{r}}^{1}dq \, q^{2}
\left(1+\frac{q^{2}+m_1m_2}{\varepsilon_{1}\varepsilon_{2}}\right)\left[1-\cos[(m_1-m_2)\tau]\right]
~~;~~ \mbox{for} ~~\frac{(m_1-m_2)\tau}{m_1m_2}\ll 1 \label{Islownrel}
\end{eqnarray}

\begin{eqnarray}
 I_f(\tau) & = &  \frac{1}{2}\int_{q_r}^{1}dq \, q^{2}
\left(1-\frac{q^2+m_1m_2}{\varepsilon_1\varepsilon_2}\right)-
\frac{(m_1-m_2)^2}{m_1m_2(m_1+m_2)~\tau}\left\{\sin\Big[(m_1+m_2)(1+\frac{1}{2m_1m_2})\tau\Big]-
\right.\nonumber \\
&  & \left. q^3_r
\sin\Big[(m_1+m_2)(1+\frac{q^2_r}{2m_1m_2})\tau\Big]\right\}+
\mathcal{O}\left(\frac{1}{\tau^2}\right) ~~;~~\mbox{for}~~
\frac{(m_1+m_2)\tau}{m_1m_2}\gg 1
\nonumber \\
I_f(\tau) & = &  \frac{1}{2}\int_{q_{r}}^{1}dq \, q^{2}
\left(1-\frac{q^{2}+m_{1}m_{2}}{\varepsilon_{1}\varepsilon_{2}}\right)\left[1-\cos[(m_1+m_2)\tau]\right]
~~;~~ \mbox{for} ~~\frac{(m_1+m_2)\tau}{m_1m_2}\ll 1
\label{Ifastnrel}
\end{eqnarray}

}

\end{itemize}

\begin{figure}[ht!]
\begin{center}
\includegraphics[height= 8cm,width=7cm,keepaspectratio=true]{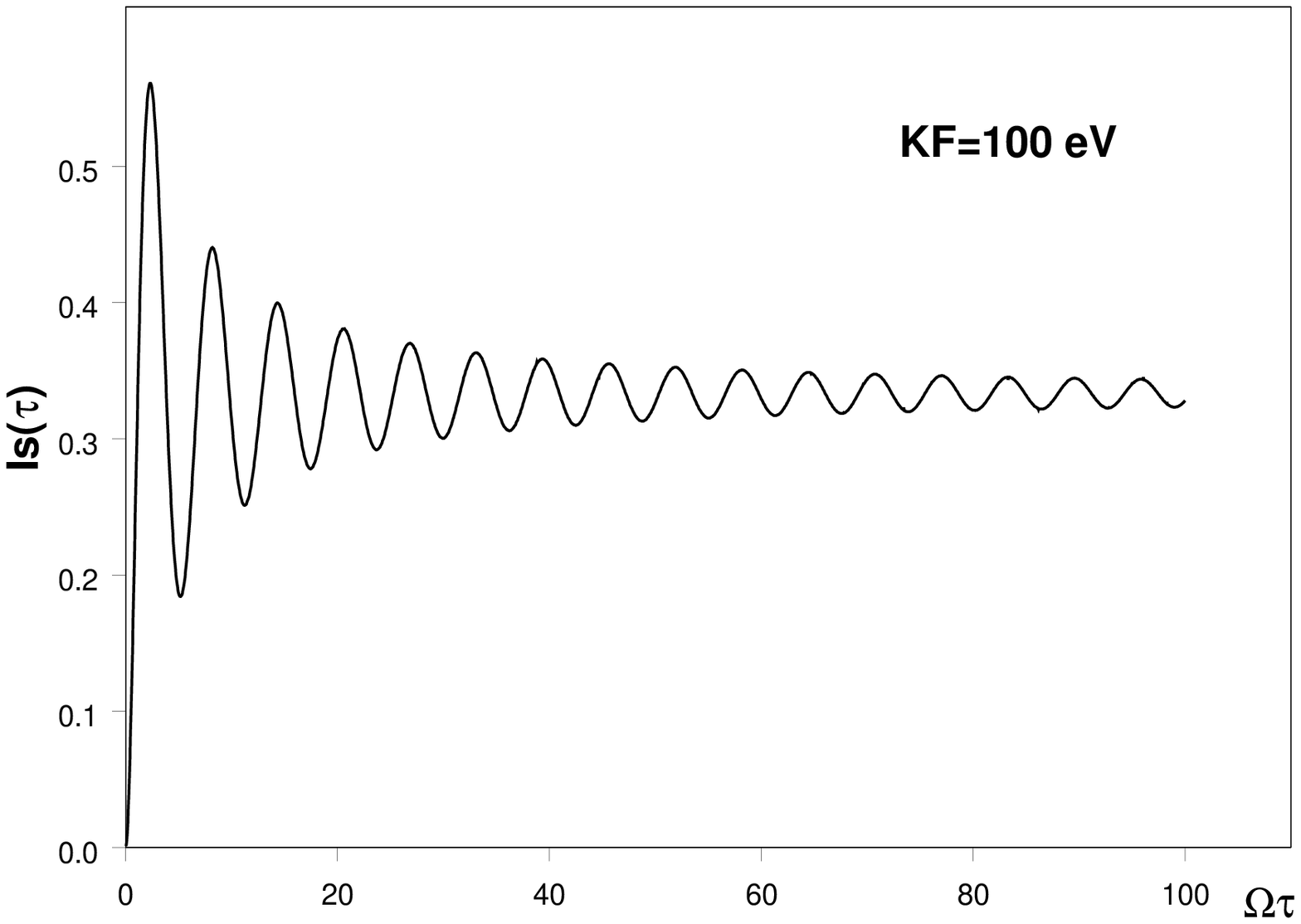}
\includegraphics[height= 8cm,width=7cm,keepaspectratio=true]{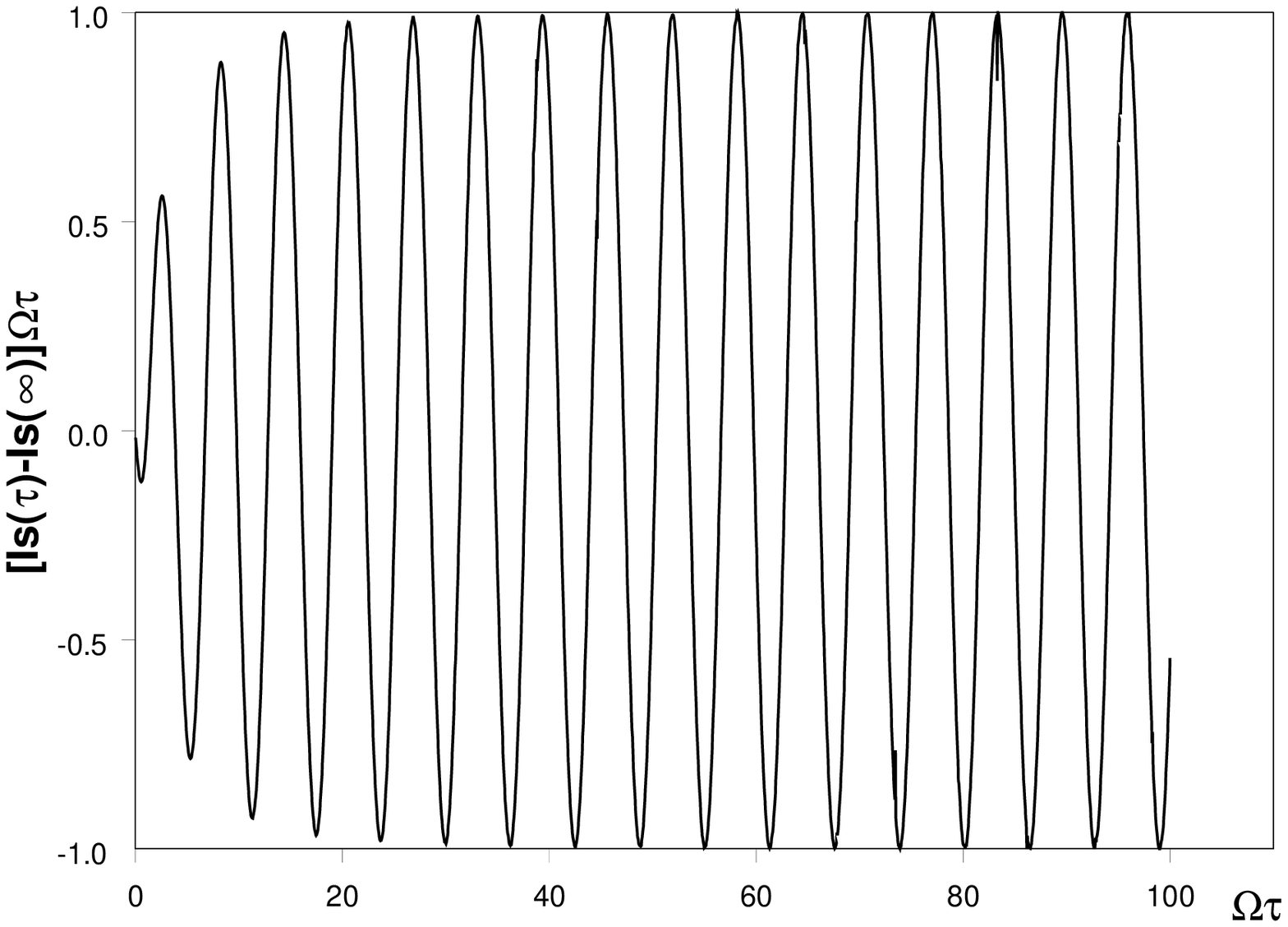}
\caption{$I_s(\tau)$ and $[I_s(\tau)-I_s(\infty)]\times(\Omega
\tau)$ for $k^e_F=100 ~eV\; ; \; k^\mu_F=0\; ; \; \bar{M}=0.25 ~
eV\; ; \; \Delta M^{2}\simeq 7\times 10^{-5}(eV)^{2}$ vs. $\Omega
\tau$, with $\Omega=\delta m^2=7\times 10^{-9}$.}
\label{fig:Islow}
\end{center}
\end{figure}

In both cases, the flavor asymmetry density  at asymptotically
long time is  given by

\begin{equation}\label{asyflavor}
\frac{1}{V}(q_e(t)-q_\mu(t)) \rightarrow \left[
\mathcal{N}^{(e)}-\mathcal{N}^{(\mu)}\right]
\cos^2(2\theta)+\mathcal{O}(1/t)
\end{equation}

The power law fall-off is a consequence of dephasing between
different flavor modes that are not Pauli blocked.
Fig.(\ref{fig:Islow}) displays the slow contribution $I_s(\tau)$
and its asymptotic limit given by eqn. (\ref{Islowrel}) in the
relativistic case.

\section{Distribution functions of neutrinos and antineutrinos}

The distribution functions are given by equations
(\ref{nel})-(\ref{muqs}) for which after lengthy but
straightforward algebra we  find the following expressions

\begin{equation}
I^{(e)}(k,t)=
n^{(e)}(k)-2n^{(e)}(k)\mathcal{A}(k,t)-\frac{k^{2}}{\omega_{e}^{2}(k)}\left[n^{(e)}(k)-\left(1-\bar{n}^{(e)}(k)\right)\right]
\mathcal{B}(k,t)
\end{equation}

\begin{eqnarray}
J^{(e)}(k,t)  &=&
\left[\frac{C^{2}S^{2}(M_{1}-M_{2})^{2}+M_{1}M_{2}+k^{2}}{\omega_{e}(k)\omega_{\mu}(k)}
\left(n^{(\mu)}(k)-[1-\bar{n}^{(\mu)}(k)]\right)
+\left(n^{(\mu)}(k)+[1-\bar{n}^{(\mu)}(k)]\right)\right]\mathcal{A}(k,t)
\nonumber \\
&&-\frac{k^{2}}{\omega_{e}(k)\omega_{\mu}(k)}\left[n^{(\mu)}(k)-\left(1-\bar{n}^{(\mu)}(k)\right)\right]\mathcal{C}(k,t)
\end{eqnarray}

\noindent where $n^{(e)}(k)$ and $\bar{n}^{(e)}(k)$ are the
initial distribution functions given by eqn. (\ref{flavnumb}) and

\begin{eqnarray}
\mathcal{A}(k,t)  & = &  C^{2}S^{2}\left[
\left(1-\frac{k^{2}+M_{1}M_{2}}{E_{1}(k)E_{2}(k)}\right)\sin^{2}\left(\frac{E_{1}(k)+E_{2}(k)}{2}~t\right)+
\right. \label{bigA} \\ & & \left.
\left(1+\frac{k^{2}+M_{1}M_{2}}{E_{1}(k)E_{2}(k)}\right)\sin^{2}\left(\frac{E_{1}(k)-E_{2}(k)}{2}~t\right)
\right] \label{BigA}\\
\mathcal{B}(k,t) & = & C^{4}S^{4}(M_{1}-M_{2})^{2}
\left[\frac{1}{E_{1}(k)}\sin(E_{1}(k)t)
-\frac{1}{E_{2}(k)}\sin(E_{2}(k)t)\right]^{2}
\label{BigB}\\
\mathcal{C}(k,t) & =   & C^{2}S^{2}(M_{1}-M_{2})^{2}
\left[\frac{C^{2}S^{2}}{E^2_{1}(k)}\sin^{2}(E_{1}(k)t)
+\frac{C^{2}S^{2}}{E^2_{2}(k)}\sin^{2}(E_{2}(k)t)\right. \nonumber \\
&& \left.
-\frac{2C^{2}S^{2}-1}{E_{1}(k)E_{2}(k)}\sin(E_{1}(k)t)\sin(E_{2}(k)t)\right].\label{BigC}
\end{eqnarray}

The expressions for $\bar{I}^{(e)}(k,t)$ and $\bar{J}^{(e)}(k,t)$
are obtained from those for ${I}^{(e)}(k,t)$ and ${J}^{(e)}(k,t)$
above by the replacement

\begin{equation}\label{antidist}
n^{(e)}(k) \longleftrightarrow [1-\bar{n}^{(e)}(k)]\; ; \;
n^{(\mu)}(k) \longleftrightarrow [1-\bar{n}^{(\mu)}(k)]
\end{equation}

Finally the expressions for
$I^{(\mu)}(k,t);\bar{I}^{(\mu)}(k,t);J^{(\mu)}(k,t);\bar{J}^{(\mu)}(k,t)$
are obtained from those for the electron neutrino by the
replacement

\begin{equation}\label{mudist}
n^{(e)}(k) \longleftrightarrow n^{(\mu)}(k)\; ; \;
\bar{n}^{(e)}(k) \longleftrightarrow \bar{n}^{(\mu)}(k) \; ; \;
\omega_e(k) \longleftrightarrow \omega_\mu(k)\; ; \; C^2
\longleftrightarrow S^2
\end{equation}

These dynamical factors
$\mathcal{A}(k,t);\mathcal{B}(k,t);\mathcal{C}(k,t)$ are
determined by the time evolution while their pre-factors in the
expressions for the distribution functions are determined by the
initial state. The dynamical factors clearly reveal again the
different time scales. Terms that feature the contributions
$e^{\pm 2 iE_{(1,2)} t};e^{\pm  i(E_1+E_2) t}$ arise from
particle-antiparticle interference and their contribution is
proportional to $\left( \Delta M^2/\bar{M}^2 \right)$ and those
that feature $e^{\pm i(E_1-E_2) t}$ arise from particle-particle
(or anti-particle- anti-particle) interference. We can find the
asymptotic distribution functions at long time by averaging the
oscillatory terms over a time scale \emph{longer than the longest}
scale $\sim \bar{M}/\Delta M^2$. This time averaging procedure
leads to

\begin{eqnarray}
\overline{\mathcal{A}(k,t)} & = & C^2 S^2 \label{Aav}\\
\overline{\mathcal{B}(k,t)} & = & \overline{\mathcal{C}(k,t)} =
\frac{1}{2}C^{4}S^{4}(M_{1}-M_{2})^{2} \left[\frac{1}{E^2_{1}(k)}+
\frac{1}{E^2_{2}(k)}\right] \label{Bav}
\end{eqnarray}

The above expressions are exact and therefore valid for any value
of the neutrino masses $M_1,M_2$. However, the most recent
compilation\cite{kamland,smirnov} of data suggests that in the two
flavor case the mass eigenstates are almost degenerate with
$\Delta M^2 \sim 7\times 10^{-5}~(eV)^2$ and the most recent
cosmological constraint from WMAP\cite{WMAP} suggests that the
average value of the mass $\bar{M}$ is $\lesssim 0.25 ~(eV)$. In
terms of the $\bar{M}$ and $\Delta M^2$ introduced in eqn.
(\ref{defmass}), we find

\begin{equation}\label{flavormasses}
m_e  =  \bar{M}\left[1+ \frac{\Delta M^2}{4
\bar{M}^2}\cos(2\theta)\right] ~;~ m_\mu  = \bar{M}\left[1-
\frac{\Delta M^2}{4 \bar{M}^2}\cos(2\theta)\right]
\end{equation}

In terms of the small ratio $\Delta M^2/\bar{M}^2 \sim 10^{-3}$ we
find the average of the distribution functions over the longest
time scale to be given by

\begin{eqnarray}
n^{(e)}_{av}(k) & = &  n^{(e)}(k) - 2 C^2 S^2
\left(n^{(e)}(k)-n^{(\mu)}(k)\right)-
\mathcal{R}[k,n^{(\alpha)},\bar{n}^{(\alpha)}] \label{nasiel}\\
\bar{n}^{(e)}_{av}(k) & = &  \bar{n}^{(e)}(k) - 2 C^2 S^2
\left(\bar{n}^{(e)}(k)-\bar{n}^{(\mu)}(k)\right)-
\mathcal{R}[k,n^{(\alpha)},\bar{n}^{(\alpha)}] \label{nasielbar}\\
n^{(\mu)}_{av}(k) & = &  n^{(\mu)}(k) + 2 C^2 S^2
\left(n^{(e)}(k)-n^{(\mu)}(k)\right)-
\mathcal{R}[k,n^{(\alpha)},\bar{n}^{(\alpha)}] \label{nasimu}\\
\bar{n}^{(\mu)}_{av}(k) & = &  \bar{n}^{(\mu)}(k) + 2 C^2 S^2
\left(\bar{n}^{(e)}(k)-\bar{n}^{(\mu)}(k)\right)-
\mathcal{R}[k,n^{(\alpha)},\bar{n}^{(\alpha)}] \label{nasimubar}
\end{eqnarray}

\noindent with

\begin{eqnarray}
\mathcal{R}[k,n^{(\alpha)},\bar{n}^{(\alpha)}]  & = &
\frac{k^2\bar{M}^2}{\bar{E}^4(k)}\left(\frac{\Delta
M^2}{4\bar{M}^2}\right)^2~C^2S^2\left[ 4C^2S^2
(n^{(e)}(k)+\bar{n}^{(e)}(k)-1)+(n^{(\mu)}(k)+\bar{n}^{(\mu)}(k)-1)
\right] \nonumber \\ & &  + \mathcal{O}\left(\left(\frac{\Delta
M^2}{4\bar{M}^2}\right)^3 \right) \label{R}\\
\bar{E}(k)= \sqrt{k^2+\bar{M}^2} \label{barE}
\end{eqnarray}

The term $\mathcal{R}[k,n^{(\alpha)},\bar{n}^{(\alpha)}] $ arises
from the overlap between particle and antiparticle spinors which
features the  small quantity $\left(\frac{\Delta
M^2}{4\bar{M}^2}\right)$.

\textbf{Flavor pair production and normal ordering: }

The expressions (\ref{nasiel}-\ref{nasimubar}) with that for the
corrections given by eqn. (\ref{R}) point out an important and
subtle aspect of the dynamics of mixing. Consider that the initial
density matrix is the flavor \emph{vacuum}, namely set
$n^{(e)}(k)=\bar{n}^{(e)}(k)=n^{(\mu)}(k)=\bar{n}^{(\mu)}(k)=0$.
The asymptotic limit of the distribution functions
(\ref{nasiel}-\ref{nasimubar}) is given to lowest non trivial
order in the  ratio $\Delta M^2/\bar{M}^2$ by

\begin{eqnarray}\label{partprod}
n^{(e)}(k,\infty)  =
\bar{n}^{(e)}(k,\infty)=n^{(\mu)}(k,\infty)=\bar{n}^{(\mu)}(k,\infty)=
&& \frac{k^2\bar{M}^2}{4\bar{E}^4(k)}\left(\frac{\Delta
M^2}{4\bar{M}^2}\right)^2
\sin^2(2\theta)(1+\sin^2(2\theta))+\nonumber
\\&& \mathcal{O}\left(\left(\frac{\Delta M^2}{4\bar{M}^2}\right)^3
\right)
\end{eqnarray}

This result clearly indicates that the time evolution results in
the creation of particle-antiparticle pairs of electron and muon
neutrinos. This is of course a consequence of the non-vanishing
overlap between positive and negative energy spinors which results
in that a destruction operator for flavor neutrinos develops a
component corresponding to a \emph{creation} operator of
antineutrinos during time evolution, and viceversa. In leading
order in the degeneracy, the typical momentum of the pair created
is $k\sim \bar{M}$ therefore these are typically low momentum
pairs of flavor neutrinos.

Furthermore a remarkable aspect of this pair production process
via neutrino mixing is that the distribution function of the
produced particles falls off very \emph{slowly} at high energies,
namely $n_{prod}(k,\infty) \propto 1/k^2$. As a result there is a
\emph{divergent} number of pairs produced as a consequence of
mixing and time evolution. Since the particles and antiparticles
are produced in pairs, the flavor charge vanishes, but the
individual distribution functions feature a contribution from the
pair production process. A normal ordering prescription must be
appended to subtract the infinite number of particles created,
however unlike normal ordering in the usual free field theory,
which subtracts a constant, in the case of mixing such normal
ordering requires a subtraction of a \emph{distribution function}.

This is a novel and subtle phenomenon, \emph{flavor pair
production} which is a direct many particle consequence of mixing
and oscillations. Since this phenomenon is a consequence of the
interference between particle and antiparticle states is
suppressed by the small quantity $(\Delta M^2/\bar{M})^2$.

Regardless of whether this phenomenon of flavor pair production
has any bearing on the cosmology and or astrophysics of neutrinos,
it is a genuine many body aspect inherent to the field theory of
neutrino mixing that deserves to be studied in its own right as a
fundamental aspect of the field theory of mixing.

 \textbf{Off-diagonal densities}: Even
when the initial density matrix is diagonal in the flavor basis
and therefore there are no off-diagonal \emph{initial}
correlations, these develop upon time evolution as a consequence
of flavor mixing. Following the same steps described above for the
distribution functions, we find the off-diagonal density to be
given by the following expression

\begin{eqnarray}
<\nu_{e}^{\dagger}(\vk,t)\nu_{\mu}(\vk,t)> &=&
-2\frac{C^{2}-S^{2}}{CS}
\left[\left(n^{(e)}(k)-\bar{n}^{(e)}(k)\right)-\left(n^{(\mu)}(k)-\bar{n}^{(\mu)}(k)\right)\right]\,\mathcal{A}(k,t)
\nonumber \\
&&+iCS\sin\left[\left(E_{1}(k)-E_{2}(k)\right)t\right] \times\nonumber \\
&&\,\,\,\,
\Bigg\{\frac{1}{\omega_{e}(k)}\left[n^{(e)}(k)-\left(1-\bar{n}^{(e)}(k)\right)\right]
\left[(E_{1}(k)+E_{2}(k))-(M_{1}-M_{2})
\left(\frac{S^{2}M_{1}}{E_{1}(k)}-\frac{C^{2}M_{2}}{E_{2}(k)}\right)\right] \nonumber \\
&&  \,\,\,\,\,\,\,\,\,\,\,
-\frac{1}{\omega_{\mu}(k)}\left[n^{(\mu)}(k)-\left(1-\bar{n}^{(\mu)}(k)\right)\right]\left[(E_{1}(k)+E_{2}(k))-(M_{1}-M_{2})
\left(\frac{C^{2}M_{1}}{E_{1}(k)}-\frac{S^{2}M_{2}}{E_{2}(k)}\right)\right]\Bigg\}
\nonumber \\
&&+iCS\sin\left[\left(E_{1}(k)+E_{2}(k)\right)t\right]\times\nonumber \\
&&\,\,\,\,
\Bigg\{\frac{1}{\omega_{e}(k)}\left[n^{(e)}(k)-\left(1-\bar{n}^{(e)}(k)\right)\right]\left[(E_{1}(k)-E_{2}(k))-(M_{1}-M_{2})
\left(\frac{S^{2}M_{1}}{E_{1}(k)}+\frac{C^{2}M_{2}}{E_{2}(k)}\right)\right] \nonumber \\
&&  \,\,\,\,\,\,\,\,\,\,\,
-\frac{1}{\omega_{\mu}(k)}\left[n^{(\mu)}(k)-\left(1-\bar{n}^{(\mu)}(k)\right)\right]\left[(E_{1}(k)-E_{2}(k))-(M_{1}-M_{2})
\left(\frac{C^{2}M_{1}}{E_{1}(k)}+\frac{S^{2}M_{2}}{E_{2}(k)}\right)\right]\Bigg\}\nonumber
\\
\end{eqnarray}

\noindent with $\mathcal{A}(k,t)$ given by eqn. (\ref{bigA}). The
expressions for the distribution functions and the off-diagonal
density can be simplified by expanding the coefficients of the
oscillatory functions up to leading order in the small quantity
$\left (\Delta M^2/\bar{M}^2\right)$. We find

\begin{equation}\label{distsimple}
n^{(e)}(k,t)= n^{(e)}(k) -
(n^{(e)}(k)-n^{(\mu)}(k))2C^2S^2\left[1-\cos[(E_1(k)-E_2(k))t]\right]+\mathcal{O}\left(\frac{\Delta
M^2}{4\bar{M}^2} \right)^2
\end{equation}

\noindent the other distribution functions may be found from the
expression above by the replacements in eqns.
(\ref{antidist},\ref{mudist}). Their time averages over the long
time scale coincides with the leading expressions in eqns.
(\ref{nasiel}-\ref{nasimubar}).  The off-diagonal density
simplifies to the following expression

\begin{eqnarray}\label{offdia}
<\nu_{e}^{\dagger}(\vk,t)\nu_{\mu}(\vk,t)> &=& -2SC
\Bigg\{2(C^2-S^2)\left[\left(n^{(e)}(k)- \bar{n}^{(e)}(k)\right)-
\left(n^{(\mu)}(k)-
\bar{n}^{(\mu)}(k)\right)\right]\sin^2\left[\left(E_1(k)-E_2(k)
\right)\frac{t}{2} \right]\nonumber\\ && -i
\left[\left(n^{(e)}(k)+ \bar{n}^{(e)}(k)\right)-
\left(n^{(\mu)}(k)+ \bar{n}^{(\mu)}(k)\right)\right]\sin
\left[\left(E_1(k)-E_2(k) \right)t \right]\Bigg\} \nonumber \\ & &
+ \mathcal{O}\left(\frac{\Delta M^2}{4\bar{M}^2} \right)^2
\end{eqnarray}

The terms of $\mathcal{O}\left(\frac{\Delta M^2}{4\bar{M}^2}
\right)^2$ again involve terms that oscillate with the sum of the
frequencies corresponding to particle-antiparticle interference as
well as terms that oscillate with the difference of the
frequencies arising from the overlap of the particle (or
antiparticle) spinor wavefunctions for different masses. The
analysis that was presented for the same type of contributions in
$I_s(\tau);I_f(\tau)$ above highlight that the
particle-antiparticle interference becomes subdominant on time
scales $t > 1/\bar{M}$. Hence the first terms
$\mathcal{O}\left(\frac{\Delta M^2}{4\bar{M}^2} \right)^0$ in the
approximations (\ref{distsimple}, \ref{offdia}) determine the
dynamics of the distribution functions and the off-diagonal
correlator in leading order in the small ratio $\frac{\Delta
M^2}{4\bar{M}^2}$ for $t>>1/\bar{M}$.

\subsection{Equilibrated gas of mass eigenstates}
Although we have focused on the case in which the initial density
matrix is diagonal in the flavor basis, for completeness we now
study the case in which the initial density matrix describes an
ensemble of \emph{mass eigenstates} in equilibrium. Therefore this
initial density matrix is diagonal in the mass basis and commutes
with the Hamiltonian. This situation thus describes a state of
equilibrium in which the occupation numbers do not evolve in time
(in the non-interacting theory). In this case we find

\begin{eqnarray}
<[\bar{\psi}_{i}(\vk,0)]_{r}[\psi_{i}(\vk,0)]_{s}> &=& n^{(i)}(k)
\left(\frac{\!\!\not\!k_{i}+M_{i}}{2E_{i}(k)}\right)_{sr}
+\left(1-\bar{n}^{(i)}(k)\right)\left[\gamma^{0}\frac{\,\,\!\!\not\!k_{i}-M_{i}}{2E_{i}(k)}\gamma^{0}\right]_{sr}
\equiv [N_{i}(\vk)]_{sr}\\
\!\!\not\!k_{i}&=& \gamma^{0}E_{i}(\vk)-\vec{\gamma}\cdot\vk
\end{eqnarray}

\noindent where $n^{(i)}(k)$ are the occupation numbers of mass
eigenstates assumed to depend only on the energy. Just as we did
in our previous analysis it proves convenient to write the above
correlator in the following form

\begin{eqnarray}
N_{i}(\vk)&=&
\,\,\!\!\not\!Q_{i}+\tilde{M}_{i} \\
\!\!\not\!Q_{i}&=&\gamma^{0}Q_{i}^{0}-\vec{\gamma}\cdot
\vec{Q}_{i}, \\
Q_{i}^{0}&=&\frac{1}{2}\left[n^{(i)}(k)+\left(1-\bar{n}^{(i)}(k)\right)\right], \\
\vec{Q}_{i}&=&\frac{\vec{k}}{2E_{i}(k)}\left[n^{(i)}(k)-\left(1-\bar{n}^{(i)}(k)\right)\right],
\\
\tilde{M}_{i}&=&\frac{M_{i}}{2E_{i}(k)}\left[n^{(i)}(k)-\left(1-\bar{n}^{(i)}(k)\right)\right].
\end{eqnarray}

Since the density matrix commutes with the full Hamiltonian, the
distribution functions of the \emph{flavor} eigenstates do not
depend on time. Following the procedure detailed above we find the
following results

\begin{eqnarray}
n^{(e)}(k)&=&\frac{C^{2}}{2}\left[\left(1+\frac{k^2+m_{e}M_{1}}{\omega_{e}(k)E_{1}(k)}\right)n^{(1)}(k)
+\left(1-\frac{k^2+m_{e}M_{1}}{\omega_{e}(k)E_{1}(k)}\right)\left(1-\bar{n}^{(1)}(k)\right)\right]
\nonumber \\
&&+\frac{S^{2}}{2}\left[\left(1+\frac{k^2+m_{e}M_{2}}{\omega_{e}(k)E_{2}(k)}\right)n^{(2)}(k)
+\left(1-\frac{k^2+m_{e}M_{2}}{\omega_{e}(k)E_{2}(k)}\right)\left(1-\bar{n}^{(2)}(k)\right)\right]
\\
\bar{n}^{(e)}(k)
&=&1-\frac{C^{2}}{2}\left[\left(1-\frac{k^2+m_{e}M_{1}}{\omega_{e}(k)E_{1}(k)}\right)n^{(1)}(k)
+\left(1+\frac{k^2+m_{e}M_{1}}{\omega_{e}(k)E_{1}(k)}\right)\left(1-\bar{n}^{(1)}(k)\right)\right]
\nonumber \\
&&-\frac{S^{2}}{2}\left[\left(1-\frac{k^2+m_{e}M_{2}}{\omega_{e}(k)E_{2}(k)}\right)n^{(2)}(k)
+\left(1+\frac{k^2+m_{e}M_{2}}{\omega_{e}(k)E_{2}(k)}\right)\left(1-\bar{n}^{(2)}(k)\right)\right]
\end{eqnarray}

Using the relations given by eqn. (\ref{flavormasses}) we find to
leading order in $\Delta M^2/\bar{M}^2$

\begin{eqnarray}\label{nemassini}
n^{(e)}(k)&=& C^{2}n^{(1)}(k)
+S^{2}n^{(2)}(k)+\mathcal{O}\left( \frac{\Delta M^2}{4\bar{M}^2}\right)^2  \\
n^{(\mu)}(k)&=& S^{2}n^{(1)}(k)
+C^{2}n^{(2)}(k)+\mathcal{O}\left( \frac{\Delta M^2}{4\bar{M}^2}\right)^2\nonumber \\
\bar{n}^{(e)}(k)&=& C^{2}\bar{n}^{(1)}(k)
+S^{2}\bar{n}^{(2)}(k)+\mathcal{O}\left( \frac{\Delta M^2}{4\bar{M}^2}\right)^2\nonumber \\
\bar{n}^{(\mu)}(k)&=& S^{2}\bar{n}^{(1)}(k)
+C^{2}\bar{n}^{(2)}(k)+\mathcal{O}\left( \frac{\Delta
M^2}{4\bar{M}^2}\right)^2
\end{eqnarray}

\section{``Effective'' (free) field theory description:}
Let us summarize the lessons learned in the analysis of the
previous section in order to establish a set of criteria with
which to develop an effective description of the dynamics in the
case in which the mass eigenstates are nearly degenerate as
confirmed by the experimental situation or in the relativistic
case.

\begin{itemize}
\item{  For nearly degenerate mass eigenstates there is a
hierarchy of scales determined by i) $k_F$ or temperature (T), ii)
the average mass $\bar{M}$ and iii) the mass difference $M_1-M_2$.
The experimental situation seems to confirm the near degeneracy
with $|M_1-M_2| \ll \bar{M}$, therefore at least two scales are
widely separated. Furthermore \emph{if} $k_F$ and or $T$
(temperature) are such that $k_F;T \gg \bar{M}$ which describes a
relativistic case, then all three scales are widely separated with
the hierarchy $k_F,T \gg \bar{M}\gg |M_1-M_2|$. The dynamics
studied above reveals all three scales. }

\item{  The time evolution of the distribution functions, flavor
asymmetry and off-diagonal correlators all feature terms that
oscillate with the frequencies  $E_1(k)+E_2(k)$, $2E_{1,2}(k)$,
and also terms which oscillate with the difference
$E_1(k)-E_2(k)$. The former arise from the interference between
particle and antiparticle states of equal or different masses and
determine the short time scales $t \lesssim 1/\bar{M}$ , while the
latter arise from interference between particle states (or
antiparticle states) of different masses and determine the long
time scales $t \gtrsim \bar{M}/\Delta M^2$. The terms that
oscillate with the fast time scales average out on these fast
scales and their coefficients are of order $\Delta M^2/\bar{M}^2$
and hence small in the nearly degenerate case. These coefficients
result from the overlap between positive and negative energy
spinors of slightly different masses.  The coefficients of the
terms that oscillate on the long time scale are of
$\mathcal{O}(1)$ and result from the overlap between positive
energy spinors  (or between negative energy spinors) of different
masses. }

\item{ The contributions to the distribution functions and
off-diagonal correlators from the terms with fast and slow
oscillations are \emph{comparable} within the short time scale
$t\lesssim 1/\bar{M}$ but for times longer than this scale the
contributions from the terms with fast oscillations are suppressed
with respect to those with slow oscillations at least by
$\mathcal{O}\left( \frac{\Delta M^2}{4\bar{M}^2 }\right)^2$.  }
\end{itemize}

We seek to obtain a description of the oscillation dynamics on
scales much larger than $1/\bar{M}$ when the contribution from the
fast oscillations have averaged out to quantities that are
proportional to powers of the small ratio $\frac{\Delta
M^2}{4\bar{M}^2}$ and can therefore be neglected in the nearly
degenerate case.

In the nearly degenerate case $\Delta M^2/\bar{M}^2 \ll 1$ the
masses $m_e,m_\mu, M_1, M_2 \sim \bar{M}$ (see eqns.
(\ref{flavormasses}),\ref{massmass})), thus in order to isolate
the leading order terms as well as to understand corrections in
the degeneracy parameter $\Delta M^2/\bar{M}^2 $ it proves
convenient to expand the positive and negative energy spinors in
terms of this small parameter. A straightforward computation  in
the standard Dirac representation of the Dirac gamma matrices
leads to the following result for the flavor positive and negative
energy spinors (see eqn. (\ref{flavor}))

\begin{eqnarray}\label{flavspinap}
U_{\vk ,\lambda}^{(\alpha)}   & =  & \left[ 1\pm \frac{\Delta
M^2}{4\bar{M}^2
}\frac{\bar{M}}{\bar{E}(k)}\cos(2\theta)\left(\frac{\gamma^0\bar{E}(k)-\bar{M}}{2\bar{E}(k)}\right)+\mathcal{O}\left(\frac{\Delta
M^2}{4\bar{M}^2 }\right)^2\right]\mathcal{U}_{\vk
,\lambda} \nonumber \\
 V_{-\vk ,\lambda}^{(\alpha)} & = & \left[ 1 \mp \frac{\Delta M^2}{4\bar{M}^2}
\frac{\bar{M}}{\bar{E}(k)}\cos(2\theta)\left(\frac{\gamma^0\bar{E}(k)+\bar{M}}{2\bar{E}(k)}\right)+\mathcal{O}\left(\frac{\Delta
M^2}{4\bar{M}^2 }\right)^2\right]\mathcal{V}_{\vk ,\lambda}
\end{eqnarray}

\noindent with

\begin{equation}
\bar{E}(k)= \sqrt{k^2+\bar{M}^2}
\end{equation}

\noindent and  the upper sign corresponds to $\alpha= {e}$ and the
lower sign to $\alpha = \mu$. The spinors $\mathcal{U}_{\vk
,\lambda}$, $\mathcal{V}_{\vk ,\lambda}$ are positive and negative
energy solutions respectively of the Dirac equation with mass
$\bar{M}$ with unit normalization. Similarly for the positive and
negative energy spinors associated with the mass eigenstates
$F^{(i)}_{\vk ,\lambda}; G^{(i)}_{-\vk ,\lambda}$ (see eqn.
(\ref{psis})), we find

\begin{eqnarray}\label{massspinap}
F_{\vk ,\lambda}^{(i)}   & =  & \left[ 1\pm \frac{\Delta
M^2}{4\bar{M}^2
}\frac{\bar{M}}{\bar{E}(k)}\left(\frac{\gamma^0\bar{E}(k)-\bar{M}}{2\bar{E}(k)}\right)+\mathcal{O}\left(\frac{\Delta
M^2}{4\bar{M}^2 }\right)^2\right]\mathcal{U}_{\vk
,\lambda} \nonumber \\
G_{-\vk ,\lambda}^{(i)} & = & \left[ 1 \mp \frac{\Delta
M^2}{4\bar{M}^2}
\frac{\bar{M}}{\bar{E}(k)}\left(\frac{\gamma^0\bar{E}(k)+\bar{M}}{2\bar{E}(k)}\right)+\mathcal{O}\left(\frac{\Delta
M^2}{4\bar{M}^2 }\right)^2\right]\mathcal{V}_{\vk ,\lambda}
\end{eqnarray}

\noindent with the same spinors $\mathcal{U}_{\vk
,\lambda};\mathcal{V}_{\vk ,\lambda}$, where the upper sign
corresponds to $i=1$ and the lower sign to $i=2$.

It is clear from the approximations (\ref{flavspinap}) and
(\ref{massspinap}) that the overlap between positive and negative
energy spinors of different masses is
$\mathcal{O}\left(\frac{\Delta M^2}{4\bar{M}^2 } \right)^2$. For
times much larger than the fast time scale, the corrections to the
spinors are subdominant and can be neglected and the fields
associated with the flavor and mass eigenstates are expanded as

\begin{eqnarray}
\nu_{\alpha}(\vk,t) & = & \sum_{\lambda}
 \left(
 \alpha_{\vk, \lambda}^{(\alpha)}(t)~\mathcal{U}_{\vk,\lambda}+
 \beta_{-\vk , \lambda}^{(\alpha) \dagger}(t)~\mathcal{V}_{-\vk , \lambda}
 \right) + \mathcal{O}\left(\frac{\Delta M^2}{4\bar{M}^2 } \right) \label{flavapx} \\
 \psi_{i}(\vk,t) &  = &  \sum_{\lambda} \left(
 a_{\vk, \lambda}^{(i)}~\mathcal{U}_{\vk, \lambda}~e^{-i E_i(k)t}+
 b_{-\vk, \lambda}^{(i) \dagger}~\mathcal{V}_{-\vk, \lambda}~e^{i E_i(k)t}
 \right)+\mathcal{O}\left(\frac{\Delta M^2}{4\bar{M}^2 } \right) . \label{massapx}
\end{eqnarray}

We can now find the  relation between the creation and
annihilation operators of flavor states and those of mass
eigenstates by using eqn. (\ref{RotationMatrix}),  to leading
order in the degeneracy parameter we find

\begin{eqnarray}
\alpha_{\vk, \lambda}^{(e)}(t) & = &  C a_{\vk,
\lambda}^{(1)}~e^{-i E_1(k)t}+ S a_{\vk, \lambda}^{(2)}~e^{-i
E_2(k)t}\label{alfaele}\\
\alpha_{\vk, \lambda}^{(\mu)}(t) & = &  C a_{\vk,
\lambda}^{(2)}~e^{-i E_2(k)t}- S a_{\vk, \lambda}^{(1)}~e^{-i
E_1(k)t}\label{alfamu}
\end{eqnarray}
\noindent where we have neglected terms of
$\mathcal{O}\left(\frac{\Delta M^2}{4\bar{M}^2 } \right)$, and
similar relations hold for the annihilation operators of the
respective antiparticles $\beta_{\vk, \lambda}^{(\alpha)}(t)$. It
is clear that the approximations leading to the relations
(\ref{alfaele}) and (\ref{alfamu}) are more generally valid not
only in the nearly degenerate case but also in the relativistic
case $k\gg M_{1,2}$ regardless of the value of the mass
difference, since in this case the common spinors are those of
massless Dirac fermions in all cases.

In this approximation, the evolution equation for the Heisenberg
operators $\alpha_{\vk, \lambda}^{(\alpha)}(t)$ does \emph{not}
follow directly from any Dirac equation, but can be obtained
straightforwardly by taking time derivatives of these operators in
eqns. (\ref{alfaele},\ref{alfamu})  and using the relations
(\ref{alfaele},\ref{alfamu}) to re-write the result in terms of
the operators themselves. In the leading order approximation
particles and antiparticles do not mix since the overlap between
the spinors $\mathcal{U}_{\vk,\lambda}$ and
$\mathcal{V}_{-\vk,\lambda}$ vanishes (in free field theory) and a
straightforward calculation leads to the following equations of
motion

\begin{equation}\label{eqofmot}
i \frac{d}{dt} \left(\begin{array}{c}
  \alpha_{\vk, \lambda}^{(e)}(t) \\
  \alpha_{\vk, \lambda}^{(\mu)}(t) \\
\end{array}\right) = \left[\bar{E}(k) \left(\begin{array}{cc}
  1 & 0 \\
  0 & 1 \\
\end{array}\right)-\Omega(k) \left(\begin{array}{cc}
  -\cos(2\theta) & \sin(2\theta) \\
  \sin(2\theta) & \cos(2\theta) \\
\end{array}\right) \right]\left(\begin{array}{c}
  \alpha_{\vk, \lambda}^{(e)}(t) \\
  \alpha_{\vk, \lambda}^{(\mu)}(t) \\
\end{array}\right)
\end{equation}

\noindent with

\begin{eqnarray}
\bar{E}(k) & = & \frac{1}{2}(E_1(k)+E_2(k))= \sqrt{k^2+\bar{M}^2}+
\mathcal{O}\left(\frac{\Delta M^2}{4\bar{M}^2 }
\right)\label{Ebar}\\ \Omega(k) & = & \frac{1}{2}(E_1(k)-E_2(k)) =
\frac{\Delta M^2}{4\bar{E}(k)} + \mathcal{O}\left(\frac{\Delta
M^2}{4\bar{M}^2 } \right)\label{Omega}
\end{eqnarray}

\noindent and a similar equation of motion for the annihilation
operators for flavor antiparticles $\beta_{\vk,
\lambda}^{(\alpha)}(t)$. These equations of motion look to be the
familiar ones for neutrino
oscillations\cite{book1}-\cite{raffelt,stodolsky,haxton,MSWI,mannheim},
but these are equations for the Heisenberg field operators, rather
than for the single particle wave-functions. Once the time
evolution of the operators is found, we can find the time
evolution of \emph{any multiparticle state}. Furthermore the
regime of validity of these equations is more general, they  are
valid \emph{either} in the nearly degenerate case $\Delta
M^2/\bar{M}^2 \ll 1 $ for any value of the momentum, or in the
relativistic limit for arbitrary value of the masses provided that
$k\gg M_1,M_2$.

Inverting the relation between the operators for flavor and mass
states at the initial time, namely writing the operators $a_{\vk,
\lambda}^{(i)}$ in terms of  $\alpha_{\vk, \lambda}^{(\alpha)}(0)$
using eqns. (\ref{alfaele}, \ref{alfamu}) at $t=0$,  we find
(again to leading order)

\begin{eqnarray}
\alpha_{\vk, \lambda}^{(e)}(t) & = & \alpha_{\vk,
\lambda}^{(e)}(0)\left[C^2 e^{-iE_1(k)t}+S^2e^{-iE_2(k)t} \right]+
SC~ \alpha_{\vk, \lambda}^{(\mu)}(0)\left[ e^{-iE_2(k)t}-
e^{-iE_1(k)t}
\right]\label{alfet} \\
\alpha_{\vk, \lambda}^{(\mu)}(t) & = & \alpha_{\vk,
\lambda}^{(\mu)}(0)\left[C^2 e^{-iE_2(k)t}+S^2e^{-iE_1(k)t}
\right]+ SC~ \alpha_{\vk, \lambda}^{(e)}(0)\left[ e^{-iE_2(k)t}-
e^{-iE_1(k)t} \right]\label{alfmut}
\end{eqnarray}

For the antiparticle operators we find the same equations with
$\alpha_{\vk, \lambda}^{(\alpha)}\rightarrow \beta_{\vk,
\lambda}^{(\alpha)}$.

The   Heisenberg field operators given by eqns.
(\ref{alfet},\ref{alfmut}) (and the equivalent for the
antiparticle operators) are the solutions of the equations of
motion (\ref{eqofmot}).

The time evolution of the distribution functions in an initial
density matrix that is diagonal in the flavor basis follows from a
straightforward calculation using the above time evolution. We
find

\begin{eqnarray}\label{dist2}
n^{(e)}(k,t) & = &  \langle \alpha_{\vk, \lambda}^{(e)\dagger}(t)
\alpha_{\vk, \lambda}^{(e)}(t) \rangle = n^{(e)}(k)-
\frac{1}{2}\sin^2(2\theta)
\left(n^{(e)}(k)-n^{(\mu)}(k)\right)\left[1-\cos[(E_1(k)-E_2(k))t]\right]
\label{neapx} \\
n^{(\mu)}(k,t) & = &  \langle \alpha_{\vk,
\lambda}^{(\mu)\dagger}(t) \alpha_{\vk, \lambda}^{(\mu)}(t)
\rangle = n^{(\mu)}(k)+ \frac{1}{2}\sin^2(2\theta)
\left(n^{(e)}(k)-n^{(\mu)}(k)\right)\left[1-\cos[(E_1(k)-E_2(k))t]\right]
\label{nmuapx}
\end{eqnarray}

The distribution functions for antiparticles to leading order is
obtained from the above results by the replacements
$n^{(\alpha)}\rightarrow \bar{n}^{(\alpha)}$. A straightforward
calculation following the above steps leads to the result

\begin{eqnarray}\label{offdia2}
<\nu_{e}^{\dagger}(\vk,t)\nu_{\mu}(\vk,t)> &=& -\sin(2\theta)
\Bigg\{2\cos(2\theta)\left[\left(n^{(e)}(k)-
\bar{n}^{(e)}(k)\right)- \left(n^{(\mu)}(k)-
\bar{n}^{(\mu)}(k)\right)\right]\sin^2\left[\left(E_1(k)-E_2(k)
\right)\frac{t}{2} \right]\nonumber\\ && -i
\left[\left(n^{(e)}(k)+ \bar{n}^{(e)}(k)\right)-
\left(n^{(\mu)}(k)+ \bar{n}^{(\mu)}(k)\right)\right]\sin
\left[\left(E_1(k)-E_2(k) \right)t \right]\Bigg\}
\end{eqnarray}

The results (\ref{dist2}) and (\ref{offdia2}) reproduce the
leading order expressions found in the previous section, eqns.
(\ref{distsimple},\ref{offdia}). Thus this ``effective'' free
field theory description reproduces the leading order results
either in the nearly degenerate case $\Delta M^2 \ll \bar{M}^2$ or
in the relativistic case. Furthermore either the effective
equations of motion (\ref{eqofmot}) or alternatively the time
evolution (\ref{alfet},\ref{alfmut}) (and those for antiparticles)
lead to a set of closed evolution equations for \emph{bilinears}.
These are most conveniently written by introducing a fiducial spin
$\overrightarrow{S}=(S_x,S_y,S_z)$ with the following components

\begin{eqnarray}\label{spin}
 S_x(\vk,\lambda;t)  & = &  i\left(\alpha_{\vk,
\lambda}^{(\mu)\dagger}(t)\alpha_{\vk,
\lambda}^{(e)}(t)-\alpha_{\vk,
\lambda}^{(e)\dagger}(t)\alpha_{\vk,
\lambda}^{(\mu)}(t)\right)\label{Sx}\\
S_y(\vk,\lambda;t)  & = &  \left(\alpha_{\vk,
\lambda}^{(\mu)\dagger}(t)\alpha_{\vk,
\lambda}^{(e)}(t)+\alpha_{\vk,
\lambda}^{(e)\dagger}(t)\alpha_{\vk,
\lambda}^{(\mu)}(t)\right)\label{Sy}\\
S_z(\vk,\lambda;t)  & = &  \left(\alpha_{\vk,
\lambda}^{(e)\dagger}(t)\alpha_{\vk,
\lambda}^{(e)}(t)-\alpha_{\vk,
\lambda}^{(\mu)\dagger}(t)\alpha_{\vk,
\lambda}^{(\mu)}(t)\right)\label{Sz}
\end{eqnarray}

\noindent and a fiducial magnetic field
$\overrightarrow{B}=(B_x,B_y,B_z)$ with components
\begin{equation}
\overrightarrow{B}(k)=2\Omega(k)(0,-\sin(2\theta),\cos(2\theta))
\end{equation}

\noindent in terms of which the equations for the bilinears are
akin to the Bloch equations for a spin $\overrightarrow{S}$
precessing in the magnetic field $\overrightarrow{B}$ namely

\begin{equation}\label{bloch}
\frac{d~\overrightarrow{S}(\vk,\lambda;t)}{dt} =
\overrightarrow{S}(\vk,\lambda;t)\times \overrightarrow{B}(k)
\end{equation}

 The antiparticle operators  obey independently a similar set of equations.
   To leading order in
$\Delta M^2/\bar{M}^2$ there is no mixing between particles and
antiparticles (suppressed by \emph{two} powers of this small
ratio), therefore the number of electron plus muon neutrinos is
conserved independently of that for antineutrinos, namely

\begin{equation}\label{conservation}
\frac{d}{dt} \left(\alpha_{\vk,
\lambda}^{(e)\dagger}(t)\alpha_{\vk,
\lambda}^{(e)}(t)+\alpha_{\vk,
\lambda}^{(\mu)\dagger}(t)\alpha_{\vk,
\lambda}^{(\mu)}(t)\right)=0
\end{equation}
\noindent and similarly for the operators $\beta_{\vk,
\lambda}^{(\alpha)}$. The set of equations above, for Heisenberg
operators is akin to the equations of motion for the ``single
particle'' density matrix  obtained in ref.\cite{samuel}, which
are  equivalent to those investigated in
refs.\cite{bell,mckellar,pantaleone2,lunar1,wong}.

In the study of synchronized
oscillations\cite{samuel,wong,bell,beacom}, a self-consistent
Hartree-Fock approximation is introduced which leads to a Bloch
equation like (\ref{bloch}) but where the magnetic field
$\overrightarrow{B}$ acquires a correction from the
self-consistent Hartree terms which arise from forward scattering
off neutrinos in the medium.

This effective formulation neglects the dynamics of flavor pair
production discussed above since such phenomenon is suppressed by
two powers of the small ratio $\Delta M^2/\bar{M}^2$.

\subsection{Propagators: non-equilibrium correlation functions}

While the set of equations of motion (\ref{eqofmot}) and
(\ref{bloch}) are reminiscent of those for the single particle
wave functions and the single particle density matrix, in fact
there is  more information in the ``effective'' free field theory
description afforded by the \emph{operator} equations
(\ref{eqofmot}) and (\ref{bloch}) combined with the field
expansion (\ref{flavapx}). In particular, inserting the solution
of the equations of motion (\ref{alfet}, \ref{alfmut}) (and the
similar ones for the antiparticles) into the expansion
(\ref{flavapx}) for the field operators allow us to obtain
\emph{any} correlation function in the free field theory at equal
\emph{or different times}. These are the building blocks of any
systematic perturbative expansion of processes of weak
interactions. In particular the Feynman propagators, which are an
essential ingredient in \emph{any} calculation that involves
neutrinos are given by

\begin{equation}
\mathcal{S}^F_{(\alpha,\alpha')}(\vec{x}-\vec{x'};t,t')= -i\int
\frac{d^3k}{(2\pi)^3}e^{i\vk\cdot (\vec x-\vec x')}~\left[\langle
\nu_{(\alpha)}(\vk,t) \bar{\nu}_{(\alpha')}(\vk,t') \rangle
\Theta(t-t')-\langle \bar{\nu}_{(\alpha')}(\vk,t')
\nu_{(\alpha)}(\vk,t)\rangle \Theta(t'-t) \right]
\end{equation}

\noindent where the expectation values are in the initial density
matrix, which is taken to be diagonal in the flavor basis in the
present discussion.

The correlation (Wightmann) functions that enter in the Feynman
propagator are found by using the leading order expansion
(\ref{flavapx}) with the time evolution of the creation and
annihilation operators given by eqns. (\ref{alfet},\ref{alfmut})
and similar ones for $\beta^{(\alpha)}_{\vk,\lambda}(t)$. With the
purpose of highlighting the fast and slow time scales in the
propagators,  it is convenient to introduce the following
functions that evolve on the slow time scale

\begin{eqnarray}
f_k(t) & = & \cos[\Omega(k)t]-i~\cos(2\theta) \sin[\Omega(k)t]
\label{foft}\\
g_k(t) & = & i~\sin(2\theta) \sin[\Omega(k)t] \label{goft}
\end{eqnarray}

\noindent in terms of which the Heisenberg creation and
annihilation operators of flavor states are written as follows

\begin{eqnarray}
\alpha_{\vk, \lambda}^{(e)}(t) & = &
e^{-i\bar{E}(k)t}\left[\alpha_{\vk, \lambda}^{(e)}(0) f_k(t)+
\alpha_{\vk, \lambda}^{(\mu)}(0)g_k(t)
\right]\label{alfetnew} \\
\alpha_{\vk, \lambda}^{(\mu)}(t) & = &
e^{-i\bar{E}(k)t}\left[\alpha_{\vk, \lambda}^{(\mu)}(0)f^*_k(t)+
 \alpha_{\vk, \lambda}^{(e)}(0)g_k(t)
\right]\label{alfmutnew}
\end{eqnarray}

\noindent and similarly for the antiparticle Heisenberg operators
$\beta_{\vk,\lambda}^{(\alpha)}(t)$.

A straighforward calculation of the Wightman functions yields the
following results

\begin{eqnarray}
\langle \nu_{(e)}(\vk,t) \bar{\nu}_{(e)}(\vk,t') \rangle & = &
 \left(\frac{\!\!\not\!k+\bar{M}}{2\bar{E}(k)}\right) e^{-i\bar{E}(k)(t-t')}
\left[(1-n^{(e)}(k))f_k(t)f^*_k(t')+(1-n^{(\mu)}(k))
g_k(t)g^*_k(t')\right]+ \nonumber \\ & & \left(
\gamma^{0}\frac{\,\,\!\!\not\!k-\bar{M}}{2\bar{E}(k)}\gamma^{0}\right)
e^{i\bar{E}(k)(t-t')} \left[\bar{n}^{(e)}(k)
f^*_k(t)f_k(t')+\bar{n}^{(\mu)}(k)
g^*_k(t)g_k(t')\right] \label{wightel}\\
\langle \bar{\nu}_{(e)}(\vk,t') \nu_{(e)}(\vk,t) \rangle & = &
 \left(\frac{\!\!\not\!k+\bar{M}}{2\bar{E}(k)}\right) e^{-i\bar{E}(k)(t-t')}
\left[n^{(e)}(k)f_k(t)f^*_k(t')+n^{(\mu)}(k)
g_k(t)g^*_k(t')\right]+ \nonumber \\ & &
\left(\gamma^{0}\frac{\,\,\!\!\not\!k-\bar{M}}{2\bar{E}(k)}\gamma^{0}\right)
e^{i\bar{E}(k)(t-t')}
\left[(1-\bar{n}^{(e)}(k))f^*_k(t)f_k(t')+(1-\bar{n}^{(\mu)}(k))
g^*_k(t)g_k(t')\right] \label{wightel2}
\end{eqnarray}

\noindent where

\begin{equation}
\!\!\not\!k \equiv \gamma^{0}\bar{E}(k)-\vec{\gamma}\cdot \vk
\end{equation}

The Wightman function for the muon neutrino is obtained from that
of the electron by the replacement
$n^{(e)}(k),\bar{n}^{(e)}(k)\rightarrow
n^{(\mu)}(k),\bar{n}^{(\mu)}(k)$ , \emph{and} $f_k
\longleftrightarrow f^*_k$ . The off-diagonal Wightman functions
are given by

\begin{eqnarray}
\langle \nu_{(\mu)}(\vk,t) \bar{\nu}_{(e)}(\vk,t') \rangle & = &
 \left(\frac{\!\!\not\!k+\bar{M}}{2\bar{E}(k)}\right) e^{-i\bar{E}(k)(t-t')}
\left[(1-n^{(\mu)}(k))f^*_k(t)g^*_k(t')+(1-n^{(e)}(k))
f^*_k(t')g_k(t)\right]+ \nonumber \\ & &\left(
\gamma^{0}\frac{\,\,\!\!\not\!k-\bar{M}}{2\bar{E}(k)}\gamma^{0}\right)
e^{i\bar{E}(k)(t-t')} \left[\bar{n}^{(\mu)}(k)
g_k(t')f_k(t)+\bar{n}^{(e)}(k)
g^*_k(t)f_k(t')\right] \label{wightmuel}\\
\langle  \bar{\nu}_{(e)}(\vk,t') \nu_{(\mu)}(\vk,t) \rangle & = &
 \left(\frac{\!\!\not\!k+\bar{M}}{2\bar{E}(k)}\right) e^{-i\bar{E}(k)(t-t')}
\left[n^{(\mu)}(k)f^*_k(t)g^*_k(t')+n^{(e)}(k)
f^*_k(t')g_k(t)\right]+ \nonumber \\ & & \left(
\gamma^{0}\frac{\,\,\!\!\not\!k-\bar{M}}{2\bar{E}(k)}\gamma^{0}\right)
e^{i\bar{E}(k)(t-t')} \left[(1-\bar{n}^{(\mu)}(k))
g_k(t')f_k(t)+(1-\bar{n}^{(e)}(k)) g^*_k(t)f_k(t')\right]
\label{wightelmu}
\end{eqnarray}

\noindent the other off-diagonal Wightmann function is obtained
from the one above by replacing $n^{(e)} \longleftrightarrow
n^{(\mu)}$ \emph{and} $f_k \longleftrightarrow f^*_k$.

We have specifically separated the ``fast'' evolution, encoded in
the exponentials $e^{i\pm \bar{E}(k)(t-t')}$ and the ``slow''
evolution encoded in the functions $f_k;g_k$ which oscillate with
the small frequency $\Omega(k)\sim  \Delta M^2/2\bar{E}(k) $. We
emphasize that the propagators above are functions not only of the
difference $(t-t')$ but also of the \emph{sum} $(t+t')$ which
reveals a truly \emph{non-equilibrium} evolution. The manifest
lack of time translational invariance reflects the fact that the
density matrix which is diagonal in the flavor representation
\emph{does not commute} with the time evolution operator.

The discussion at the beginning of this section points out that
these  propagators are valid on time scales $t,t' \gg 1/\bar{M}$,
for which the corrections arising from the interference between
particle and antiparticle can be neglected. Therefore the
correlation functions obtained from the effective field theory
must be understood as being averaged over the fast time scales and
their validity is restricted to slow time scales.

The \emph{free field theory} propagators obtained above provide
the main ingredients to carry out a study of the weak interactions
in a neutrino background in a loop expansion.

\section{Conclusions and discussions:}
Our focus  was to study the evolution of a dense and or hot gas of
flavor neutrinos as a consequence of oscillations and mixing. The
goal was to establish an understanding of the dynamics directly
from the underlying quantum field theory, beginning with the
simplest case of free field theory and restricted to the two
flavor case.

Such study leads to a deeper understanding of the various
approximations invoked in the literature as well as recognizing
the potential corrections. Even at the level of free field theory,
which must be the starting point of any program to study the
physics of oscillations and mixing in the weak interactions, this
study reveals a wealth of dynamical phenomena that has not been
explored before within the context of neutrino oscillations in a
medium with neutrinos at finite density and temperature.

The most salient aspects of our study are the following:

\begin{itemize}
\item{A hierarchy of time scales emerges associated with different
interference phenomena. Oscillations on fast time scales $t <
1/\bar{M}$ are associated with the interference between particles
and antiparticles while oscillations on slow time scales $t>
\bar{M}/\Delta M^2$ arise from the interference between particle
(or antiparticle) states with different masses. Observationally
the situation for two flavors is that of near degeneracy, which
entails that these time scales are widely separated. Furthermore
in the relativistic limit with typical energy $\bar{E} \gg
M_1,M_2$ there is an even shorter time scale $t \sim 1/\bar{E}$. }

\item{The terms that oscillate on fast scales feature coefficients
that are determined by the overlap of positive and negative
frequency wave functions of different masses. In the relativistic
limit or in the case of near degeneracy as suggested by the recent
observations, these terms are of order $(\Delta M^2 /\bar{M}^2)^2
\sim 10^{-6} $ (or smaller in the relativistic case), while the
coefficients of terms that oscillate on the slow scales are of
$\mathcal{O}(1)$ in terms of this ratio. During the short time
scales both contributions are comparable, but for $t>>1/\bar{M}$
the contribution from the overlap between particle and
antiparticle states becomes subdominant being at least a factor
$(\Delta M^2 /\bar{M}^2)^2 \sim 10^{-6} $ smaller than the
oscillations on the slow time scale. For the values of $\bar{M}$
consistent with the recent bounds\cite{WMAP} the scale for fast
oscillations is $\sim 10^{-15}s$ these are clearly too fast for
relevant processes during BBN or neutrino processes in
astrophysics, but may be relevant for early universe cosmology. Of
course this possibility requires further and deeper studies.  }

\item{An initial flavor asymmetry relaxes to equilibrium via
dephasing between modes that are not Pauli blocked with a power
law $1/t$ on slow time scales $t > k_F/\Delta m^2$ in the
relativistic case $k_F>> \bar{M}$. We have obtained exact as well
as approximate expressions for the time evolution of the
distribution functions and off diagonal densities and discussed
their asymptotic behavior, all of which display Pauli blocking
between different flavors (see eqns.
(\ref{nasiel}-\ref{nasimubar}). For completeness we have also
studied the case of an equilibrated gas of mass eigenstates which
describes a situation of equilibrium in absence of interactions.
The non-equilibrium oscillation dynamics leads to the production
of particle-antiparticle pairs of flavored neutrinos with typical
momenta $k\sim \bar{M}$. Since this phenomenon is a direct
consequence of the overlap between particle and antiparticle
states the pair yield is suppressed by the factor $(\Delta
M^2/\bar{M}^2)^2$. }

\item{ The wide separation between the different time scales
allows to describe the dynamics on the longer time scales in terms
of an ``effective'' theory. In this effective description the
Heisenberg creation and annihilation field operators for flavor
neutrinos and antineutrinos obey the familiar Bloch type equations
and the spinor structure is common to both flavors as well as the
mass eigenstates. This effective description allows to obtain in a
simple manner the dynamics of the distribution functions, off
diagonal correlation functions and the \emph{non-equilibrium}
propagators, all of which must be understood as an average over
the fast time scales and valid only on the slow scales. }

\end{itemize}

While we have focused on the evolution of a gas of flavor
neutrinos as an \emph{initial value problem} we have not discussed
how the initial state is ``prepared''. This is an important aspect
of the physics of neutrino mixing and the weak interactions, since
weak interactions only produce flavor states the initial state (or
density matrix) must be ``prepared'' by weak interaction processes
that occur on time scales much shorter than those in which such
state will relax either via collisions or by oscillations. Clearly
we have nothing to say yet on this aspect which deserves a
thorough study.

Another aspect that deserves attention is that of the corrections
to the ``effective'' theory described above. These corrections
entail powers of the ratios that are small either in the nearly
degenerate case or in the relativistic limit. In perturbation
theory in the weak interactions, these ``small'' corrections could
conceivably be comparable to perturbative corrections in $G_F$ the
Fermi coupling, in which case the terms neglected in the effective
theory must be kept on the same footing as the contributions in
the weak coupling in the perturbative expansion. Clearly such
possibility must be evaluated  for the particular situation under
consideration.

While we have focused on the dynamics in free field theory, the
results will likely be valid in the interacting case in the case
of a low density neutrino gas (or low temperatures). Under these
circumstances the corrections to the evolution equations
associated with forward scattering off the neutrino background
(mean field), which is of order $G_F$ would be much smaller than
$\Delta M^2/\bar{M}$ and the free field theory results for the
evolution of the asymmetry may very well be valid. Furthermore,
the weak interactions only affect the left handed neutrinos but
not the right handed neutrinos which will oscillate as in a free
field theory. The mass term will then entangle the oscillations of
the right and left handed components. Such a process will be
suppressed in the relativistic limit but may introduce yet another
scale. The intriguing phenomenon of flavor pair production, a many
body feature intrinsic to the field theory of neutrino mixing and
oscillations. While it is not clear to the authors whether such
phenomenon could have potential bearing in cosmology and
astrophysics, it certainly is part of the fundamental aspects of
neutrino mixing and oscillations and deserves further study.

We are currently studying these and other possible scenarios
including interactions.

Having understood the regime of validity of the effective ``long
time'' theory as well as having obtained the necessary
non-equilibrium propagators we expect to address the issue of the
propagation of neutrinos in a dense and or hot medium, including a
neutrino background  including not  only forward scattering but
also collisional
processes\cite{dolgov,raffelt,sigl2,mckellar,enqvist} by
implementing the methods of non-equilibrium quantum field
theory\cite{noneq}.

\begin{acknowledgements}
D.B.\ thanks the  National Science Foundation for support under
grant award PHY-0242134. He also thanks S. Reddy, A. Leibovich, V.
Paolone, D. Naples for enjoyable and illuminating conversations.
\end{acknowledgements}


\begin{thebibliography}{99}
\bibitem{book1} C. W. Kim and A. Pevsner, \textit{Neutrinos in Physics and
Astrophysics}, (Harwood Academic Publishers, USA, 1993).
\bibitem{book2} R. N. Mohapatra and P. B. Pal, \textit{Massive Neutrinos in Physics and
Astrophysics}, (World Scientific, Singapore, 1998).
\bibitem{book3} M. Fukugita and T. Yanagida, \textit{Physics of Neutrinos and Applications to
Astrophysics}, (Springer-Verlag Berlin Heidelberg 2003).
\bibitem{raffelt} G. G. Raffelt, \textit{Stars as Laboratories for Fundamental
Physics}, (The University of Chicago Press, Chicago, 1996).
\bibitem{pantaleone} T. K. Kuo and J. Pantaleone, Rev. of Mod.
Phys. \textbf{61}, 937 (89).
\bibitem{dolgov} A. D. Dolgov, Surveys High Energ.Phys. {\bf 17}, 91,
(2002); Phys.Rept. \textbf{370}, 333 (2002); Nuovo Cim.
\textbf{117 B}, 1081, (2003); hep-ph/0109155 ( talk at XV
Rencontres de Physique de La Vallee d'Aoste, March, 2001); A.D.
Dolgov, S.H. Hansen, S. Pastor, S.T. Petcov, G.G. Raffelt, D.V.
Semikoz, Nucl.Phys. B632 (2002) 363-382.
\bibitem{pontecorvo} B. Pontecorvo, Zh. Eksp. Toer. Fiz.
\textbf{34},247 [Sov. Phys. JETP \textbf{7},172 (1958)]; Zh. Eksp.
Toer. Fiz. \textbf{53}, 1717 [Sov. Phys. JETP \textbf{26}, 984
(1968)]; Z. Maki, M. Nakazawa and S. Sakata, Prog. Theor. Phys.
\textbf{28}, 870 (1962).
\bibitem{bilponte} S. M. Bilenky and B. Pontecorvo, Phys. Rept.
\textbf{41}, 225 (1978).
\bibitem{giunti}  C. Giunti and M. Lavede, hep-ph/0310238; S.M. Bilenky, C. Giunti, J.A. Grifols, E.
Masso, Phys.Rept. \textbf{379}, 69 (2003); C. Giunti, Found. Phys.
Lett. \textbf{17}, 103 (2004); S. M. Bilenky and C. Giunti, Int.
J. Mod. Phys. \textbf{A16}, 3931 (2001).
\bibitem{smirnov} A. Yu. Smirnov, hep-ph/0311259; hep-ph/0306075 ;
hep-ph/0305106; hep-ph/0305106.
\bibitem{bilenky}  S. M. Bilenky, hep-ph/0402153;  W.M. Alberico, S.M.
Bilenky, hep-ph/0306239;  S. M. Bilenky, hep-ph/0210128
\bibitem{haxton}  W. C. Haxton, nucl-th/9901076, Wick C. Haxton, Barry R.
Holstein,  Am.J.Phys. \textbf{68}, 15 (2000).
\bibitem{grimus} W. Grimus, hep-ph/0307149.
\bibitem{roulet} E. Roulet, astro-ph/0011570.
\bibitem{Beuthe} M. Beuthe, Phys.Rept. \textbf{375}, 105 (2003).
\bibitem{MSWI} L. Wolfenstein, Phys. Rev. \textbf{D17}, 2369
(1978); Phys. Rev. \textbf{D20}, 2634 (1979); Phys. Lett.
\textbf{B194} 197 (1987).
\bibitem{MSWII} S. P. Mikheyev and A. Yu Smirnov, , Yad.
Fiz.\textbf{42}, 1441 (1985) (Sov. J. Nucl. Phys. \textbf{42}, 913
(1985)); Nuovo Cimento \textbf{C9}, 17 (1986); Zh. Eksp. Toer.
Fiz. \textbf{91}, 7 (1986) (Sov. Phys. JETP \textbf{64}, 4
(1986)).
\bibitem{kirilova} D. P. Kirilova and  M. V. Chizhov,
Phys.Rev. \textbf{D58} 073004 (1998); Phys.Lett. \textbf{B393} 375
(1997); hep-ph/9704269; Nucl.Phys. \textbf{B534} 447 (1998);
Nucl.Phys.Proc.Suppl. \textbf{100 } 360 (2001).
\bibitem{prakash}  M. Prakash, J. M. Lattimer, R. F. Sawyer, R. R.
Volkas, Ann.Rev.Nucl.Part.Sci. \textbf{51}, 295 (2001); S. Reddy,
M. Prakash, J. M. Lattimer, Phys.Rev. \textbf{D58}, 013009 (1998)
\bibitem{reddy} M. Prakash, J.M. Lattimer, J.A. Pons, A.W. Steiner, S.
Reddy,  Lect.Notes Phys. \textbf{578}, 364 (2001),  S. Reddy, M.
Prakash, nucl-th/9508009.
\bibitem{yakovlev}  D. G. Yakovlev, A. D. Kaminker, O. Y. Gnedin, P.
Haensel, Phys.Rept. \textbf{354}, 1 (2001).
\bibitem{stodolsky} A. Dolgov, Sov. J. Nucl. Phys. \textbf{33},700
(1981); R. A. Harris and L. Stodolsky, Phys. Lett. \textbf{116B},
464 (1982); L. Stodolsky, Phys. Rev. \textbf{D36}, 2273 (1987).
\bibitem{mannheim} P. Mannheim, Phys. Rev. \textbf{D37}, 1935
(1988).
\bibitem{sigl2} G. Raffelt, G. Sigl and L. Stodolsky, Phys. Rev.
\textbf{D45}, 1782 (1992); Phys. Rev. Lett. \textbf{70}, 2363
(1993); G. Sigl and G. Raffelt, Nucl. Phys. \textbf{B 406}, 423
(1993).
\bibitem{enqvist} K. Enqvist, K. Kainulainen and J. Maalampi,
Nucl. Phys. \textbf{B349}, 754, (1991).
\bibitem{mckellar} B. H. J. McKellar and M. J. Thompson, Phys.
Rev. \textbf{D49}, 2710 (1994).
\bibitem{barbieri} R. Barbieri and A. Dolgov, Nucl. Phys. \textbf{B
349}, 743 (1991).
\bibitem{samuel} V. A. Kostelecky, J. Pantaleone and S. Samuel,
Phys. Lett. \textbf{B315}, 46 (1993); V. Kostelecky and S. Samuel,
Phys. Lett. \textbf{B318}, 127 (1993); V. Kostelecky and S.
Samuel, Phys. Rev.  \textbf{D49}, 1740 (1994);  V. Kostelecky and
S. Samuel, Phys. Rev.  \textbf{D52}, 3184 (1995).
\bibitem{lunar1} C. Lunardini and A. Y. Smirnov, Phys. Rev.
\textbf{D64}, 073006 (2001).
\bibitem{wong} Y. Y. Y. Wong, Phys. Rev. \textbf{D66}, 025015
(2002); AIP Conf.Proc. \textbf{655}, 240 (2003).
\bibitem{beacom} K. N. Abazajian, J. F. Beacom and N. F. Bell,
Phys. Rev. \textbf{D 66}, 013008 (2002).
\bibitem{pantaleone2} J. Pantaleone, Phys. Lett. \textbf{342}, 250
(1995);  J. Pantaleone, Phys.Rev.\textbf{D46}, 510 (1992) .
\bibitem{fuller} Y. Z. Qian and G. M. Fuller, Phys. Rev.
\textbf{D51}, 1479 (1995).
\bibitem{sigl3} G.Sigl, Phys.Rev.\textbf{ D51}, 4035 (1995).
\bibitem{bell} N. F. Bell, R. R. Volkas, Y. Y. Y. Wong, Phys. Rev.
\textbf{D59}, 113001 (1999), N. F. Bell, R. Foot, R. R. Volkas,
Phys.Rev. \textbf{D58}, 105010 (1998); N. F. Bell, hep-ph/0311283.
\bibitem{notzold} D. Notzold and G. Raffelt, Nucl. Phys.
\textbf{B307}, 924 (1988).
\bibitem{nieves} J. F. Nieves, Phys. Rev. \textbf{D40}, 866
(1989).
\bibitem{dolivo} J. C. D'Olivo and J. F. Nieves, Int. Jour. of
Mod. Phys. \textbf{A11}, 141 (1996).
\bibitem{lunardini} A. Friedland and C. Lunardini, Phys. Rev.
\textbf{68}, 013007 (2003).
\bibitem{blasone} M. Blasone and G. Vitiello, Phys.Rev. \textbf{D60}, 111302 (1999);
 M. Blasone, G. Vitiello, Annals of Physics \textbf{244},283
 (1995); Erratum-ibid. \textbf{249}, 363 (1996); E. Alfinito, M. Blasone, A. Iorio, G.
 Vitiello, Phys.Lett. \textbf{B362}, 91 (1995); E.Alfinito, M.Blasone, A.Iorio, G.Vitiello,  Acta Phys.Polon.\textbf{ B27 }, 1493
 (1996); M.Blasone, P.A.Henning, G.Vitiello, hep-ph/9605335; M. Blasone, P. P. Pacheco, H. W. C. Tseung, Phys.Rev.
 \textbf{D67}073011(2003),  and
 references therein.
 \bibitem{fujii} K. Fujii, C. Habe and T. Yabuki, Phys. Rev.
 \textbf{D64}, 013011 (2001), Phys. Rev. \textbf{D59}, 113003
 (1999); K. Fujii and T. Shimomura, hep-ph/0402274.
 \bibitem{ji} C-R. Ji and Y. Mischchenko, Phys. Rev. \textbf{D65},
 096015 (2002); hep-ph/0403073.
 \bibitem{giunti2} C. Giunti, hep-ph/0312256, hep-ph/0402217; J. H. Fried,
 hep-ph/0301231.
 \bibitem{boyboltz} D. Boyanovsky, I.D. Lawrie, D.S. Lee, Phys.Rev. \textbf{D54}, 4013
 (1996);  S. M. Alamoudi, D. Boyanovsky, H. J. de Vega, R. Holman,  Phys.Rev. \textbf{D59}, 025003 (1999);
  D. Boyanovsky, H.J. de Vega, S.-Y. Wang, Phys.Rev. \textbf{D61}, 065006 (2000).
  \bibitem{kamland} K. Eguchi \emph{et.al.} (KamLAND
  collaboration), Phys. Rev. Lett.\textbf{90}, 021802 (2003).
  \bibitem{WMAP} D. N. Spergel \emph{et. al.}  (WMAP
  collaboration), Astrophys.J.Suppl. \textbf{148 }, 175 (2003).
\bibitem{noneq} D. Boyanovsky, M. D'Attanasio, H.J. de Vega, R. Holman,Phys.Rev. \textbf{D54} 1748
(1996);  D. Boyanovsky, M. D'attanasio, H.J. de Vega, R. Holman,
Phys.Rev. \textbf{D52}6805 (1995); D. Boyanovsky, M. D'Attanasio,
H. J. de Vega and R. Holman, in \textit{Third Paris Cosmology
Colloquium}, Eds. H. J. de Vega and N. Sanchez, (World Scientific,
Singapore, 1996); D. Boyanovsky, H. J. de Vega and R. Holman, in
\textit{International School of Astrophysics D. Chanlonge, 5th
course: Current topics in Astrofundamentel Physics}, Eds. N.
Sanchez and A. Zichichi, (World Scientific, Singapore, 1997).





\end{thebibliography}
\end{document}